\documentclass{ieeeaccess}
\usepackage{cite}
\usepackage{amsmath,amssymb,amsfonts}
\usepackage{algorithmic}
\usepackage{graphicx}
\usepackage{textcomp}
\usepackage{url} 
\usepackage{hyperref}

\def\BibTeX{{\rm B\kern-.05em{\sc i\kern-.025em b}\kern-.08em
    T\kern-.1667em\lower.7ex\hbox{E}\kern-.125emX}}
\begin{document}
\history{Received November 20, 2019, accepted December 21, 2019. Date of publication January 10, 2020, Current version January 15, 2020.}
\doi{10.1109/ACCESS.2020.2965547}

\title{Interacting spreading processes in multilayer networks: a systematic review}
\author{\uppercase{P. Br\'odka}\authorrefmark{1, 2}, 
\uppercase{K. Musial\authorrefmark{2}, and J. Jankowski}\authorrefmark{3}
}
\address[1]{Wroc\l{}aw University of Science and Technology, Department of Computational Intelligence, Wybrze\.ze Wyspia\'nskiego 27, Wroc\l{}aw,58-533, Poland}
\address[2]{University of Technology Sydney, Advanced Analytics Institute, School of Computer Science, 15 Broadway, Sydney, NSW 2007, Australia}
\address[3]{West Pomeranian University of Technology, Department of Computer Science and Information Technology, Zolnierska 49, Szczecin, 71-210, Poland}

\tfootnote{This work was partially supported by Polish National Science Centre, decisions no. 2016/21/D/ST6/02408 and 2016/21/B/HS4/01562, and the European Union's Horizon 2020 research and innovation programme under the Marie Sk\l{}odowska-Curie grant agreement No. 691152 (RENOIR); the Polish Ministry of Science and Higher Education fund for supporting internationally co-financed projects in 2016-2019 (agreement no. 3628/H2020/2016/2) and by Australian Research Council, grant no. DP190101087: "Dynamics and Control of Complex Social Networks"}

\markboth
{Author \headeretal: Preparation of Papers for IEEE TRANSACTIONS and JOURNALS}
{Author \headeretal: Preparation of Papers for IEEE TRANSACTIONS and JOURNALS}

\corresp{Corresponding author: Piotr Br\'odka (e-mail: piotr.brodka@pwr.edu.pl).}

\begin{abstract}
The world of network science is fascinating and filled with complex phenomena that we aspire to understand. One of them is the dynamics of spreading processes over complex networked structures. Building the knowledge-base in the field where we can face more than one spreading process propagating over a network that has more than one layer is a challenging task, as the complexity comes both from the environment in which the spread happens and from characteristics and interplay of spreads' propagation.
As this cross-disciplinary field bringing together computer science, network science, biology and physics has rapidly grown over the last decade, there is a need to comprehensively review the current state-of-the-art and offer to the research community a roadmap that helps to organise the future research in this area. Thus, this survey is a first attempt to present the current landscape of the multi-processes spread over multilayer networks and to suggest the potential ways forward.
\end{abstract}


\titlepgskip=-15pt

\maketitle

\section{Introduction}
\label{sec:introduction}

Dynamical processes over complex networks cover a variety of phenomena from phase transitions and synchronisation in networks, through walking and searching on networks, to epidemics spread and collective behaviour covering social influence, rumour and information spread as well as opinion formation \cite{barrat2008dynamical}, \cite{newman2018networks}, \cite{newman2011structure}. Spread over the networks, its characteristics and dynamics were always one of those research avenues that attracted a lot of attention \cite{nopr16}. Epidemiology was the area where first attempts to understand the spread were made and the first spread models, such as SIS or SIR, were developed~\cite{kermack1927contribution}. The predominant reason for that was the huge impact the spread of epidemics has on the connected society. Understanding how the contagion propagates in the population is crucial from the perspective of our lives and development of effective preventive measures. The consequences of epidemics in the modern, connected world can be very serious and we continuously get reports about new outbreaks \cite{epidemicnews2018}. So imagine the situation where we can clearly predict any epidemic before it occurs. This would mean that we are able to eliminate epidemics all together. Thus, there is a wealth of work done in the direction of understanding disease propagation and variety of computer science approaches were developed in this space \cite{newman2002spread}, \cite{nopr16}, \cite{waan15}. 

Epidemiology, although an important area where spread phenomenon is considered, is not the only one. In recent years, with the development of online world that led to the availability of huge social data, we gained more understanding about the rumours \cite{doerr2012rumors}, (dis-)information spread \cite{guille2013information} and how the opinions are formed \cite{watts2007influentials}. Also, analysis of spread in financial networks in the context of cascades triggered by some initial shocks and robustness of the system has recently attracted a lot of attention~\cite{gai2010contagion}, \cite{kanno2015assessing}, \cite{barro2010credit}. 
Another field, where spread analysis over a network is widely investigated are the computer networks \cite{shin2014cascading} and infrastructure networks in general~\cite{crucitti2004model}. Also, cybersecurity is a very popular area where researchers aim at understanding how the computer viruses and malware spread through computer networks \cite{van2009virus}, \cite{wang2018node}. All these areas, next to epidemics, became research fields on their own where analysing propagation characteristics and its dynamics is of pivotal importance to comprehensive understanding of both human and systems behaviour.  

Spread analysis is a cross-disciplinary field that has grown over the last few decades and is now strongly established in the computer science community \cite{meng2018securing}, \cite{van2009virus}, \cite{guille2013information}, \cite{wang2018node}, \cite{kempe2003maximizing}, \cite{cebe2018network}. If you search Scopus database for papers including spreading processes and networks you will find that among 2,093 papers 1,455 of them are from the computer science subject area\footnote{The Scopus query used "KEY ( spreading ) AND KEY ( network )", as of 30/09/2019}. One of the main reasons for that is the computational complexity of the spread modelling that cannot be tackled by any other research domain apart from computer science. Both simulation approaches, together with data-driven techniques from computer science, are the key ways to model the spread over networks.

There are two main components when it comes to the spread analysis over networked systems. One is the model of spread and another is structure over which the propagation takes place. Plethora of spread models exists such as susceptible-infected (SI) susceptible-infected-susceptible (SIS), susceptible-infected-recovered (SIR) or threshold-based models and they were widely studied and surveyed before~\cite{daley2001epidemic},~\cite{nopr16},~\cite{shakarian2015independent},~\cite{dietz1967epidemics},~\cite{valler2012spreading},~\cite{sash15} together with theoretical analysis behind them including mean field theory, Markov chains and other approaches ~\cite{zhang2016dynamics}.

When it comes to network models used in the spread modelling the focus is on three main models: (i)~Barab\'{a}si-Albert model for the scale-free network~\cite{barabasi1999emergence}, (ii)~Watts-Strogatz small-world model for the small-world network~\cite{watts1998collective}, and (iii)~Erd\H{o}s-R\'{e}nyi model for the random graph network~\cite{solomonoff1951connectivity}, \cite{erdos1959random}, \cite{erdos1960evolution}, \cite{erdHos1961strength}. For the review of those and other network models please see \cite{goldenberg2010survey} and \cite{newman2018networks}
Since it was already done in many review and research papers, we will not cover those models in detail in this work. Instead, if needed, we would like to refer the readers to the abovementioned literature to gain a better understanding of the basic spread and network models. 

The landscape of the research in spreading processes over networks can be divided into four groups as presented in Figure~\ref{fig:spread_landscape}. There are two main elements that contribute to the complexity of the analysis, i.e. (i) the number of spreading processes that are analysed and (ii) the structure over which the diffusion happens. Thus, to group the existing research, we use two dimensions, naming (i) Spread Complexity and (ii) Network Complexity expressed by a number of spreads propagating and a number of layers in the network respectively.

\Figure[t!](topskip=0pt, botskip=0pt, midskip=0pt)[width=3in]{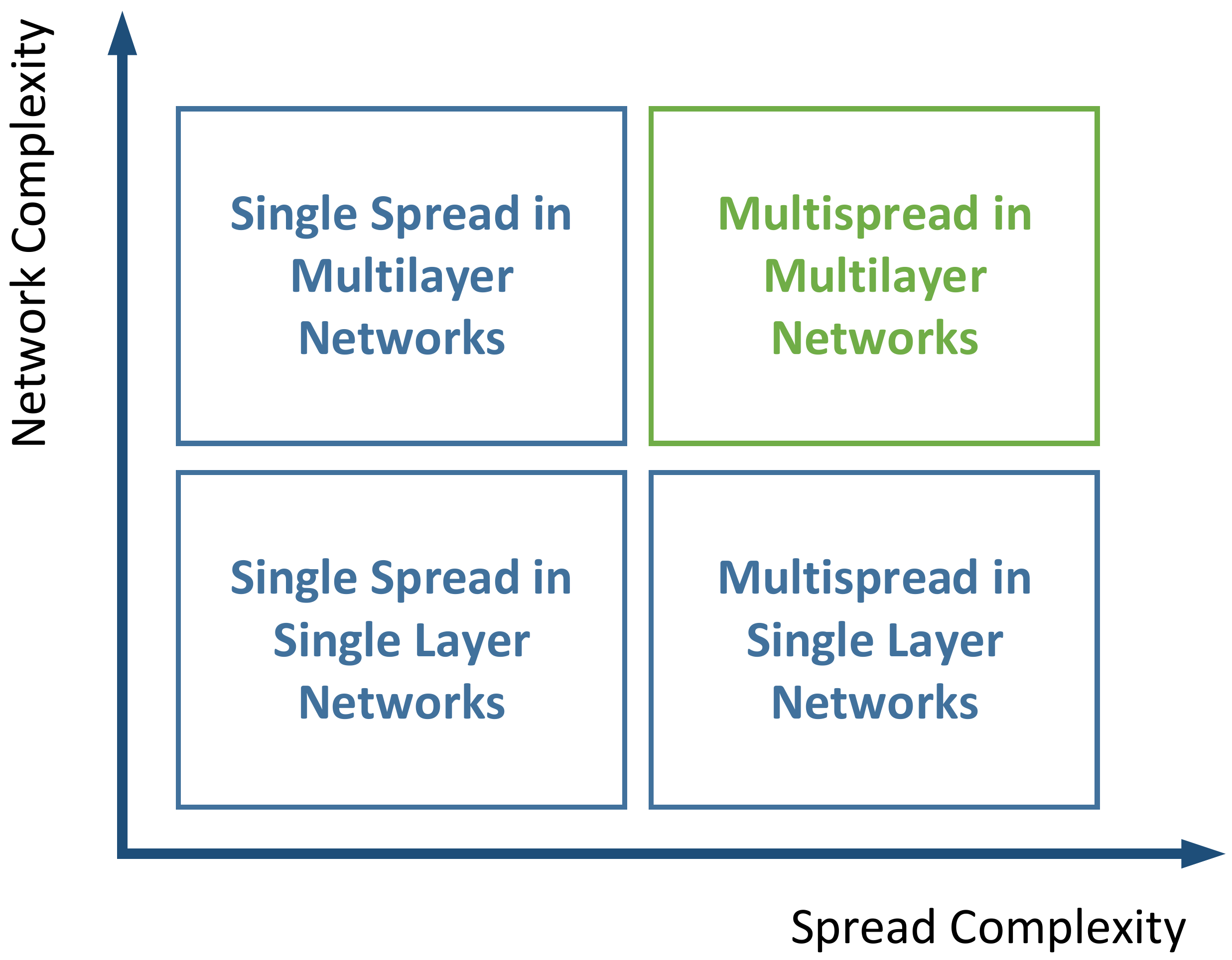}
{Landscape of research in the area of spreading processes in networks. \label{fig:spread_landscape}}


To be able to analyse such complex phenomenon, at first researchers used formalism where one spread propagated over a network describing one type of relationship between nodes (bottom, left corner of the Figure~\ref{fig:spread_landscape}). The efforts focused mainly on how disease spreads in populations \cite{fusa10}. Another avenues, embedded in the computer science community, that developed later on, are the information spread \cite{guille2013information} and rumour propagation \cite{kwon2013prominent}. 

Although the concept may seem to be simple and the field is well researched, the problem is far from being trivial with many challenges arising as we gain more understanding in this area \cite{fuba15}.

In the next phase of research, the community tried to understand how the system behaviour's changes if we include two or more spreads in the one layer network, e.g. \cite{karrer2011competing}, \cite{funk2009spread}, \cite{budak2011limiting} (bottom, right corner of the Figure~\ref{fig:spread_landscape}). Academics and practitioners looked into, e.g. how one disease can strengthen/weaken the impact of another one \cite{newman2005threshold} or how the disease can be inhibited by information \cite{zhan2018coupling} and information-driven vaccination \cite{ruan2012epidemic}. An effort was also made to analyse, e.g. how different opinions influence/compete with each other \cite{burghardt2016competing} or how the spread of truthful information can help to overcome the propagation of misinformation/gossip \cite{pan2018effective}. Another studies investigated competing viruses and ideas on fair-play networks \cite{prakash2012winner}, viral marketing performance for multiple products \cite{datta2010viral}, competitive influence in a social network \cite{wu2015maximizing}, competing opinions over evolving social networks \cite{koprulu2019battle} and mechanics of competing information in a group-based population \cite{fu2019analysis}. This gave insight into the area of competition/cooperation in the context of propagation processes, but the limiting factor was that all spreads happen through the same network. 

Research into multispreading processes over a one layer networks enabled to increase the complexity in modelling the propagation phenomenon but left the structure over which the spread happens relatively simple. This was the natural extension -- to look into single spread but over much more complex, multilayer networks (top, left corner of the Figure~\ref{fig:spread_landscape}). Each layer in multilayer structure represents one type of relationship, e.g. one layer can be the physical contact layer and another online contact layer. Different layers can also denote different types of relations, e.g. friendship on one layer and family ties on another~\cite{coba13}. Additional complexity is brought into the equation if different types of relationships (layers) are weighted depending on how close they are with higher weights assigned to closer relationships~\cite{sun2014epidemic}.
One of the first attempts to look at a single spread that could propagate over many layers was done in \cite{coba13}. From then on, the field rapidly expanded \cite{dogr16}, \cite{sash15}, covering many different, seemingly not connected, fields like epidemiology \cite{coba13}, financial markets \cite{d2014modeling}, \cite{poledna2015multi}, \cite{silva2016modeling}, games \cite{jankowski2016picture} or social media \cite{erlandsson2017seed}. 
Single spread over multilayer structure can be interpreted as a special case of multispread over multilayer network where contagions on different layers are of the same type and have the same parameters. 

So, the very much needed next step to complete the picture and create a bigger whole is to research multispread over multilayer network. We presented that in the top, right corner of the Figure~\ref{fig:spread_landscape}. This is the ultimate case that enables to consider the complexity resulting from both the propagation process and the structure over which it spreads. Multispread over multilayer networked structure is a relatively new research direction that due to its high complexity, which is brought into the equation by both heterogeneity of the multirelational networks and non-linear dynamics of the spread of multiple processes, is still in its infancy. 

\begin{figure}[!t]
\centering
\includegraphics[width=0.45\textwidth]{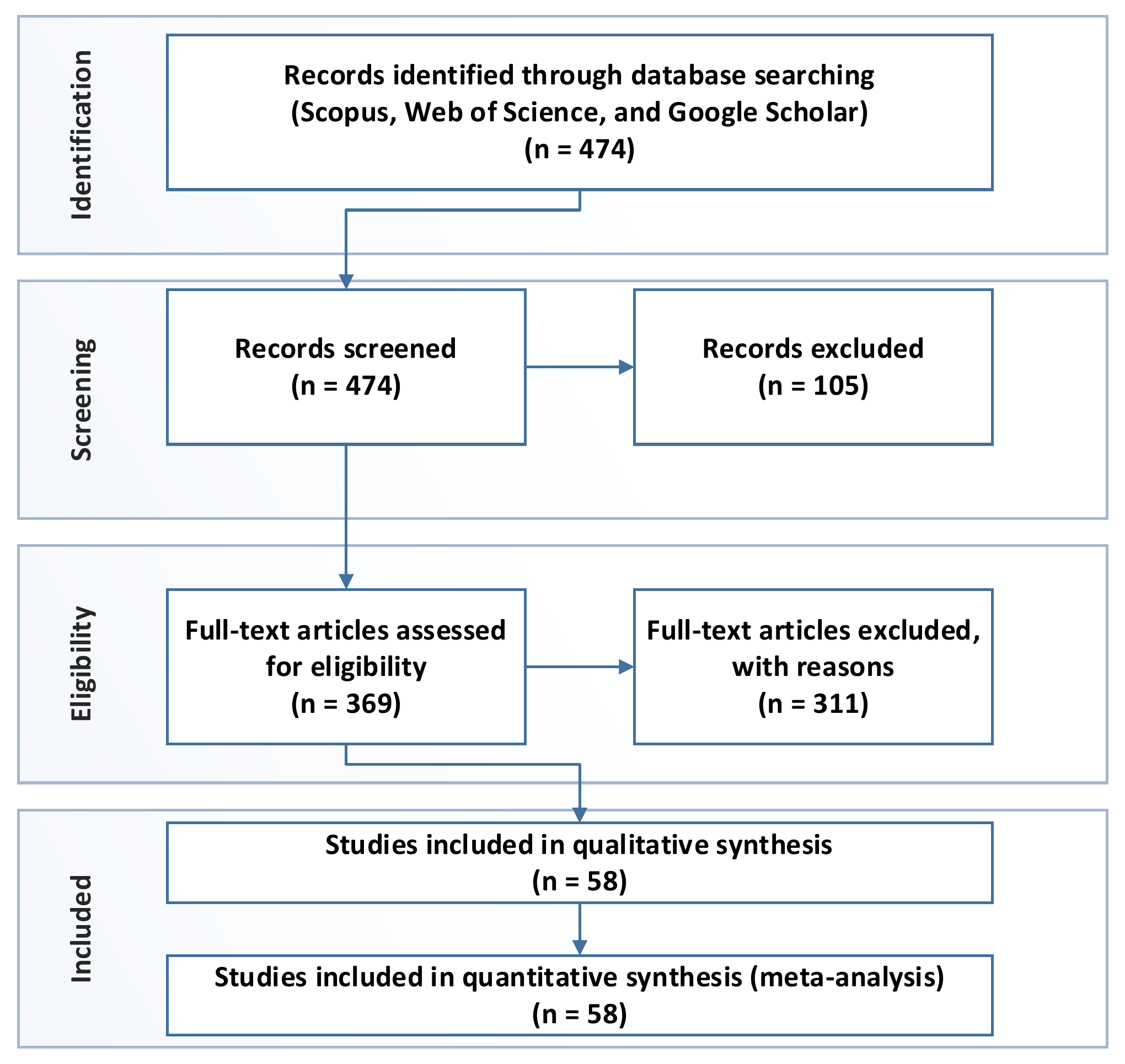}
\caption{Process of the searching for the most relevant literature for the systematic review.}
\label{fig:literature_search}
\end{figure}

First attempts to investigate multiple spreads in multilayer networked environment were done around 2006 where a spread of immunisation competed with the disease and the model used multiplex network \cite{JJA_jo2006immunization}. From then, there were 0--2 papers a year till 2013--2014 when the field started growing and the number of research outputs rapidly increased showing more and more interest in developing this research direction (see Figure~\ref{fig:numberofpaperspublished}). It is not surprising as understanding this complex phenomenon is pivotal to building a proper knowledge about how such critical processes as disease, awareness, immunisation, (mis-)information, gossip, opinion, or behaviour spread in societies. Societies that in the face of digital transformation develop more and more diverse and complex structures of interactions. 

The goal of this study is a critical and comprehensive review of existing research in the area of multiple spreading processes in multilayer networks, having two or more layers. As a result of the analysis, we present challenges arising from the limitations of the current approaches. Those challenges guided and enabled us to develop a road map that shows future directions in this exciting field of study.

The rest of our manuscript is structured as follows. First, in section~\ref{sec:approachToLiterature} the approach to the literature review is presented. We employed the Preferred Reporting Items for Systematic Reviews and Meta-Analyses methodology~\cite{moher2015preferred} to select the most relevant research papers. After that, we present the basic statistics about the chosen papers and how the research landscape changed over the last 10 years. In section \ref{sec:criticalAnalysis}, to help the reader to understand various aspects of interacting/multiple spreading processes in multilayer networks, we have decided to ask four fundamental questions - what?, where?, how? and why?. Step by step we explore the existing research from the following perspectives: (i) what spreads and (ii) where (in what type of network), (iii) how individual spreading processes and the interaction between them are modelled, and finally (iv) why the spread happens in a way it happens. After that, in section~\ref{sec:RoadMapGuidelines}, we synthesize the areas on which the future research should focus to progress the work in the area of multispread over multilayer networks. In the final section~\ref{conclusions}, we sum up our work and offer the final summary.

\section{Approach to Literature Review}
\label{sec:approachToLiterature}

\begin{figure}[!t]
\centering
\includegraphics[width=3.5in]{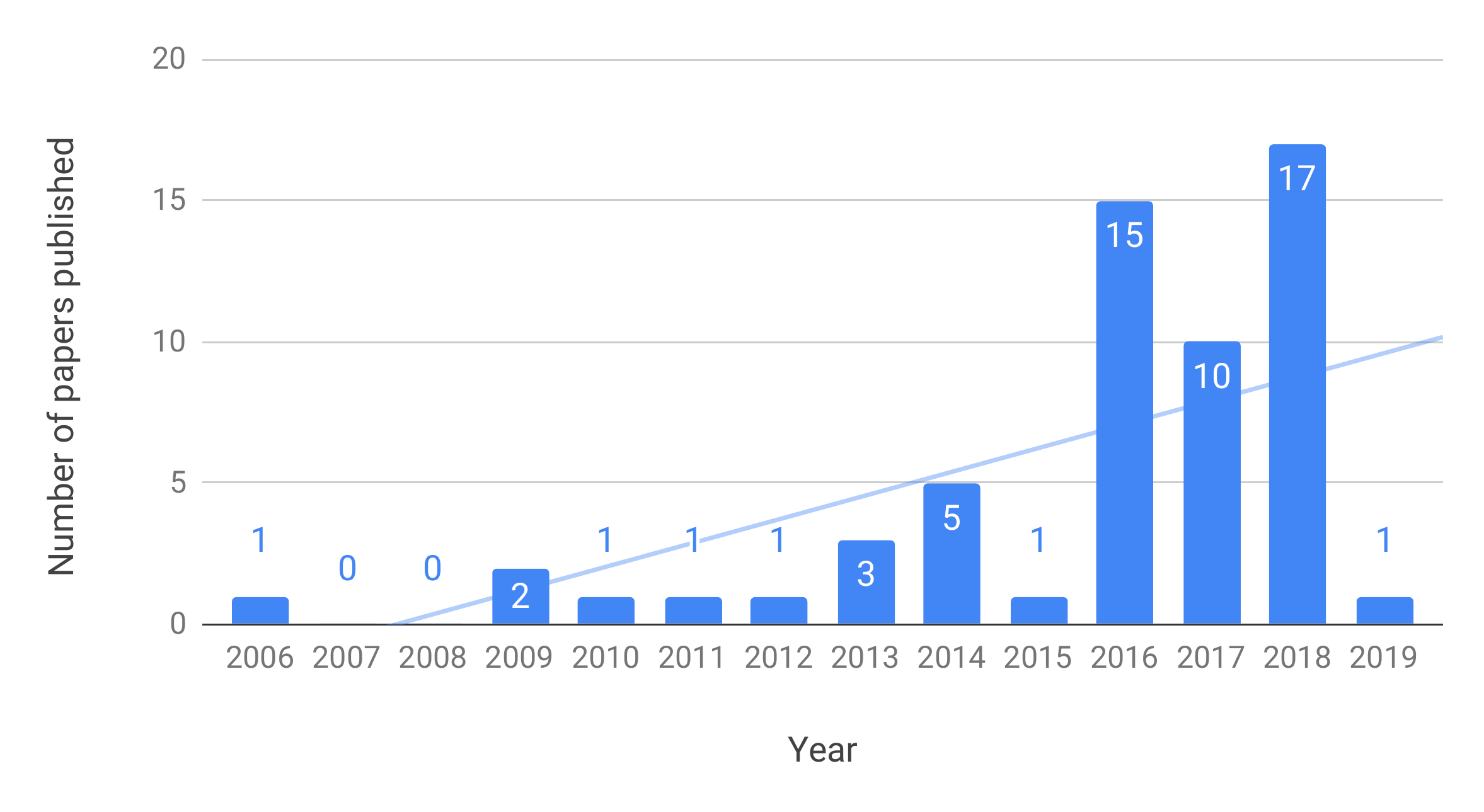}
\caption{Number of papers published per year.}
\label{fig:numberofpaperspublished}
\end{figure}

Creating comprehensive literature review starts with broad search of the relevant research. To achieve best possible result and to be able to consistently search the existing publications we decided to adopt Preferred Reporting Items for Systematic Reviews and Meta-Analyses (PRISMA) methodology~\cite{moher2015preferred} for meta-analysis. Our search for literature can be summarised in the Figure~\ref{fig:literature_search} where we present the numbers of reviewed publication and the filtering process that we followed.


During the search for relevant sources, we used the most popular search engines: Scopus, Web of Science, and Google Scholar and started with search for multispreading processes over multilayer networks.
We started from more generic keywords, to make sure that we do not oversee any research. We used the following set of keywords (i) to describe the spread: spread, propagation, diffusion, spreading, propagation, diffusion processes, multi(-)spread, multiple spread, competing, interacting, supporting, suppressing spreading processes, disease/epidemic/information/ behaviour/opinion/meme/gossip/fake news spread, (ii) to describe the networks: multi(-)layer, multi(-)dimensional, multiplex, multi(-)relational, complex networks, networked systems. We also used the combination of words describing the spreading process and the network structure to grasp all possible cases. 

The search through the databases gave us well over 400 papers, and after the initial screening, we had 369 papers that qualified to the eligibility check. The initial screening excluded papers where based on the title alone we were able to say that they do not fall in the "multispread in multilayer network" category.

During the eligibility test we discarded further 311 papers that fell outside the "multispread in multilayer network" category, but this we did by looking through the abstract and the main text of the paper. 
For each paper that passed the eligibility check we checked its references (past-cross-check) and papers that cited a given paper (future-cross-check) to see if any of those papers qualify to be included in the final meta-analysis. After that, we reached 58 publications that we included in the final review. See Figure~\ref{fig:numberofpaperspublished} for the number of relevant selected papers published each year since 2006. As mentioned before, we see a growing interest in the field of multispread over multilayer networks that is vivid when we look at a growing trend of a number of papers annually published in this space. Please note that in our analysis below, in some cases, it may look that there are more papers than 58. This is because, if one paper considers few scenarios or cases it may be counted few times e.g. in Table \ref{tab:NetworkData} we have 5 papers doing experiments on real data and 56 on synthetic data which give us 63 papers in total, however five papers \cite{weva13}, \cite{wei2012competing} \cite{scata2018quantifying}, \cite{guo2016role} and \cite{JJ8_wang2017epidemic} are counted twice since authors use both real and synthetic data in their experiments.

\begin{figure*}[!t]
\centering
\includegraphics[width=\textwidth]{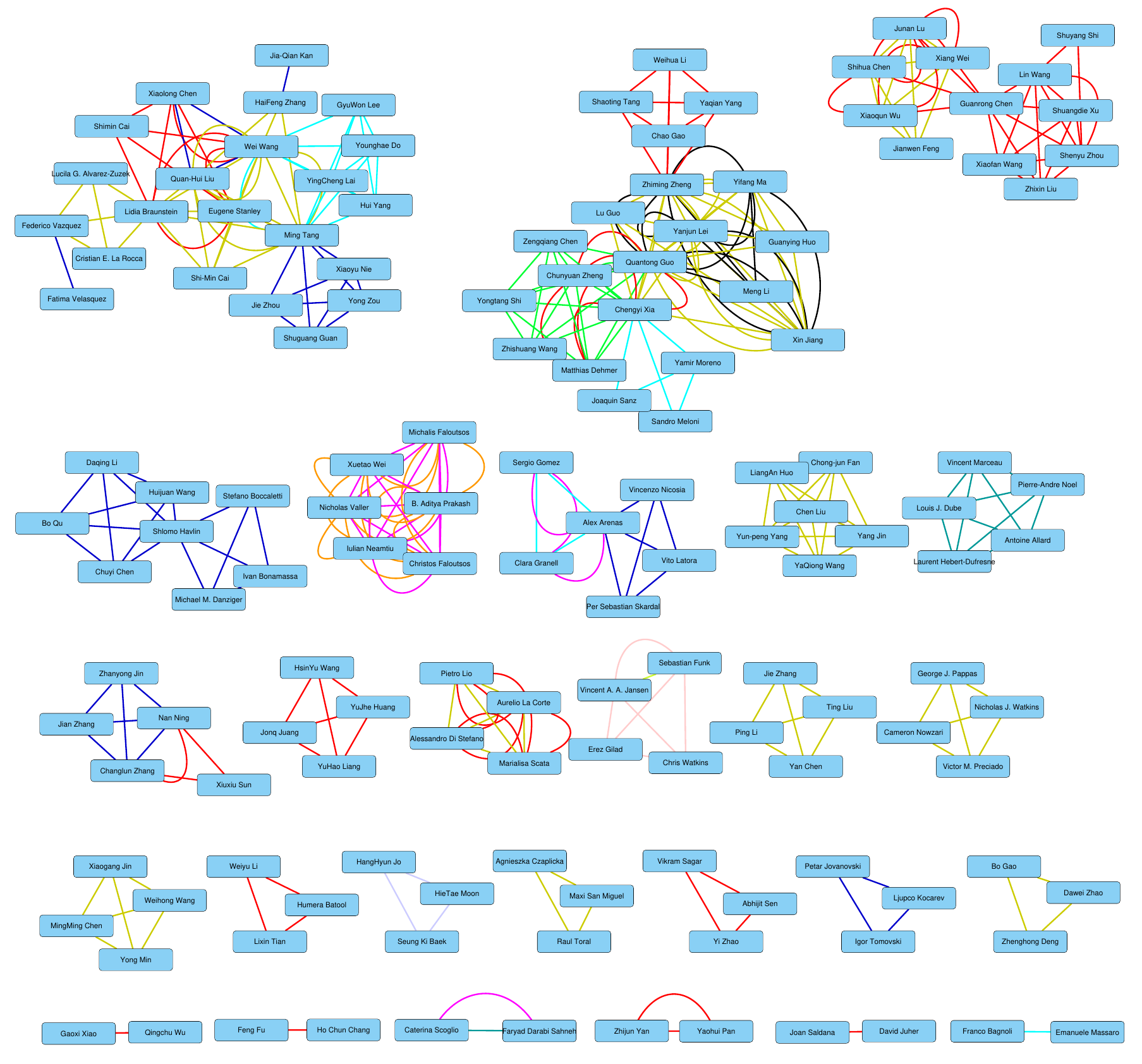}
\includegraphics[width=\textwidth]{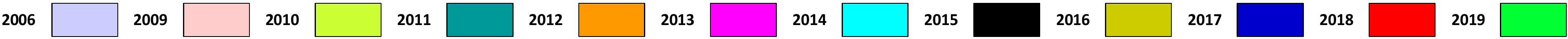}
\caption{Co-authorship network of reviewed papers. Edge colour indicates publication year of the paper.}
\label{fig:coauthors}
\end{figure*}

When we look at the authors of the reviewed papers and their co-authorship networks, the picture shows clearly how the field has developed and evolved since the first paper was published in 2006, see Figure~\ref{fig:coauthors}. 

There are two groups that consistently work on the topic of multispread in multilayer networks for the last five years. Other than that, we have several small groups that have started working in the field only recently, publishing one or two papers over the past couple of years. This shows the rapidly growing interest in the field that is also clearly visible in the Figure~\ref{fig:numberofpaperspublished}. In addition, few groups can be spotted that conducted some research when the area first appeared on the research map (2006--2010) but then discontinued their work. 

Looking at how quickly this research theme expanded over the last decade, we can anticipate that this field will attract even more attention in the following years. Furthermore, looking at the plethora of challenges identified and described in Section~\ref{sec:RoadMapGuidelines}, there is no doubt that those coming years will be full of exciting developments.

\section{CRITICAL ANALYSIS of what, where, how and why}
\label{sec:criticalAnalysis}
The key element in any literature review, next to selecting all relevant sources, is to decide how to organise the existing knowledge in a meaningful way that would enable us to tell the whole story about the current developments in a given field.
 To assist the reader in understanding various aspects of interacting/multiple spreading processes in multilayer networks we have decided to ask four fundamental questions - \textit{what?}, \textit{where?}, \textit{how?} and \textit{why?}.

\textit{What spreads?} -- describes the phenomena/medium, like virus, awareness, opinion or meme, that propagates over the network. 
\textit{Where it spreads?} -- denotes the environment and multilayer networks, with their features and topologies, on which the spreading processes are interacting. 
\textit{How it spreads?} -- indicates developed and employed spreading models together with their characteristics as well as provides information about the types of interactions between spreading processes. 
\textit{Why it spreads in that way?} -- tries to answer fundamental questions about why things happen in the way they happen. We are looking here at various aspects and features of both spreading and multilayer networks which affects the behaviour of interacting spreading processes.

Revolving the discussion around those four questions enabled us to identify drawbacks of the current approaches that, in turn, gave the foundations for defining the future research directions in this fascinating field of multispread in mulitlayer networks.

After this brief introduction, we want to invite you to read the story about how beautiful the complexity of diffusion processes over the heterogeneous networked structures is. So, let us begin the journey and immerse ourselves in the world of networks.

\subsection{What spreads?}
\label{sec:what}
Answering the question about \textit{what spreads?} sets the scene for the rest of our review. Analysing what researchers focus on in terms of what processes spread over the complex networks enables us to understand the landscape of the research in the field of multispread over multilayer structures. 

\begin{table*}[ht]
\begin{center}
\begin{tabular}{l|cccccccc}
\hline
 & virus & awareness & meme & opinion & synchronisation & social contagion\\ \hline
virus & \textbf{23.33\%} & \textbf{61.67\% }& 0.00\% &\textbf{ 3.33\% }& 0.00\% & 0.00\% \\
meme & 0.00\% & 0.00\% & \textbf{3.33\%} & 0.00\% & 0.00\% & 0.00\% \\ 
opinion & 0.00\% & 0.00\% & 0.00\% & \textbf{1.67\%} & 0.00\% & 0.00\% \\ 
decision making & 0.00\% & 0.00\% & 0.00\% &\textbf{ 1.67\%} & 0.00\% & 0.00\% \\
nutrition & 0.00\% & 0.00\% & 0.00\% & 0.00\% &\textbf{ 1.67\%} & 0.00\% \\ 
green behaviour & 0.00\% & 0.00\% & 0.00\% &\textbf{ 1.67\%} & 0.00\% & 0.00\% \\ 
social contagion & 0.00\% & 0.00\% & 0.00\% & 0.00\% & 0.00\% & \textbf{1.67\%}\\ \hline
 
\end{tabular}
\end{center}
\caption{What spreads? -- \% of reviewed literature where given two processes spread on two layer network}
\label{tab:what}
\end{table*}

\begin{table*}
\begin{center}
\begin{tabular}{l|l|p{13cm}}
\hline
\multicolumn{2}{l|}{What spreads on each layer?} & References \\ \hline
virus & virus & \cite{sahneh2014competitive} \cite{joto17} \cite{dabo17} \cite{wano16} \cite{zhxu18} \cite{wuxi18} \cite{marceau2011modeling} \cite{sanz2014dynamics} \cite{sahneh2013may} \cite{azimi2016cooperative} \cite{JJ2_wang2016structural} \cite{JJ4_wei2016unified} \cite{JJ13_wei2018cooperative} \cite{JJ29_zhou2018immunizations} \\ \hline
virus & awareness& \cite{funk2009spread}\cite{guo2016role} \cite{JJ8_wang2017epidemic}\cite{marceau2011modeling} \cite{grgo14}  \cite{jusa18} \cite{faji16} \cite{gule16} \cite{lich17} \cite{liwa16} \cite{kazh17} \cite{funk2010interacting} \cite{grgo13} \cite{czaplicka2016competition} \cite{wang2014asymmetrically} \cite{velasquez2017interacting} \cite{wang2016suppressing} \cite{scata2016impact} \cite{nicosia2017collective} \cite{nie2017impact} \cite{pan2018impact} \cite{pan2018impact2} \cite{xia2019new} \cite{zhang2017epidemic} \cite{liu2016community} \cite{zhang2018new} \cite{yang2016impact} \cite{JJ3_gao2016competing} \cite{JJ15_gao2018dynamical} \cite{JJ20_huang2018global} \cite{JJ21_zang2018effects} \cite{JJ26_zhou2017numerical} \cite{JJ28_massaro2014epidemic} \cite{JJ30_chen2018optimal} \cite{JJ32_zheng2018interplay} \cite{JJ33_sagar2018effect} \cite{JJ35_guo2015two}
 \\ \hline
virus & opinion& \cite{velasquez2017interacting} \cite{JJE_eames2009networks}\\ \hline
meme & meme & \cite{weva13} \cite{wei2012competing} \\ \hline
opinion & opinion & \cite{czaplicka2016competition} \\ \hline
opinion & green behaviour& \cite{liti18}\\ \hline
opinion & decision making & \cite{alla16} \\ \hline
nutrition & synchronisation & \cite{nicosia2017collective}\\ \hline
social contagion & social contagion & \cite{JJ34_chang2018co} \\ \hline
 
\end{tabular}
\end{center}
\caption{What spreads? -- references to the literature where given two processes spread on 2--layered network.}
\label{tab:what_ref}
\end{table*}

In the real-world, there are many situations in which we see spreading phenomena in action, from the social and behavioural perspective where the propagation of information, opinion, or certain behaviour spreads to epidemiological cases of disease, virus and/or awareness spread. But it is fair to say that the most critical and vastly discussed phenomenon in the literature on multispread over multilayer networks is the spread of multiple diseases or disease vs awareness scenarios. It covers 85\% of all reviewed studies where two processes spread over two layer networks (note that this constitutes 95\% of all literature that investigates multispread over multilayer networks). Out of this 85\%, almost 62\% are the studies where virus/disease compete with awareness/information, and over 23\% constitutes literature where two viruses/diseases interact with each other.

Another group, although much smaller, constitutes opinion/meme spread where we can have two memes \cite{weva13} and \cite{wei2012competing} or opinion \cite{czaplicka2016competition} spreading over the multilayer network. Other variations are where on one layer spreads opinion and on another (i) virus \cite{velasquez2017interacting} \cite{JJE_eames2009networks}, (ii) decision making process \cite{alla16}, or (iii) adoption of green behaviour \cite{liti18}.
This whole group is represented by 11.67\% of all reviewed literature.

In their paper Vel\'squez-Rojas and Vazquez \cite{velasquez2017interacting} present unique approach to model coupled opinion-disease system, where if two individuals have the same opinion the disease spreads with certain probability. However, if they have different opinions the probability of infection is much lower. This represents lower chance of contact between people with different opinions. Similarly, if both nodes are in the same disease state (both are Susceptible or Infected), the probability that they change their opinion is $1$. However, if the nodes are in different disease states the probability is lower than $1$, which represent lower chance of a sick (healthy) person to contact and influence the opinion of a healthy (sick) person. 

In \cite{JJE_eames2009networks}, authors show how and to what level the decision about the vaccination can be affected social influences. The opinion about the effectiveness of vaccination, influences decision about whether to vaccinate or not, and this influences spread of the disease. A similar concept is presented in \cite{liti18} where spread of the negative opinion about green behaviour influences the adoption level of such behaviour. Another similar approach, where the process of the opinion formation and spread within the society influences the decision making process among the officials, is the focus of research presented in \cite{alla16}.

Quite unique approach is presented in \cite{czaplicka2016competition}, where the authors simulate spreading of the same opinion with two competitive mechanisms: (i) threshold model (complex adoption process) and SIS model (simple adoption process). 

Some studies investigate the spread of other phenomena. For example in \cite{nicosia2017collective}, authors investigate a system "where neural dynamics and nutrient transport are bidirectionally coupled in such a way that the allocation of the transport process at one layer depends on the degree of synchronization at the other layer, and vice versa", i.e., more nutritions (food) is supplied by transport system the faster is neuron synchronization, less nutritions is available in transport system the slower the synchronization. In \cite{JJ34_chang2018co} authors analyse social contagion which explains types of collective behaviour through social contact in the areas of marketing, innovation diffusion, medicine, rumour, information spreading, emotion and others. 

For exact percentages regarding what phenomena are analysed in different studies please see Table~\ref{tab:what}, and for list of associated references see Table~\ref{tab:what_ref}.

Out of all reviewed papers, only 5\% looks at more than two layer networks. however, also in those cases, although more than two processes are considered, authors analyse spread and interactions between multiple viruses \cite{joto17}, \cite{JJ2_wang2016structural} or virus--awareness situation \cite{scata2018quantifying} where on each out of three layers, both virus and awareness spread. So those studies contribute to the biggest group of virus/virus or virus/awareness spread.

To gain a better understanding of the environment in which those different processes are implemented and how they interact with each other we need to answer questions \textit{where?} and \textit{how?}. To find answers to those please see Sections~\ref{sec:where}~and~\ref{sec:how} .

\subsection{Where it spreads?}
\label{sec:where}
The second important element of the entire process is \textit{where} the multiple processes are spreading. As outlined in the introduction, we focus on a structure that is now commonly known as a multilayer network \cite{kivela2014multilayer}, \cite{dickison2016multilayer}. In this section, we consider several characteristics of the network and the environment in which the propagation takes place. To enable comprehensive review, which takes into account all elements considered by different authors, we split our analysis into following sections (i) network topology, (ii) number of layers, (iii) multiplex vs multilayer approach, (iv) existence of edges between layers, and (v) type of potential external influence.

\subsubsection{Network Topology} \label{subsec:netTopology}
\begin{table*}
\begin{center}
\begin{tabular}{l|rp{13cm}}
\hline 
Network data & No. of papers & References \\ \hline
Real data & 5 (8\%) & \cite{weva13} \cite{wei2012competing} \cite{scata2018quantifying} \cite{JJ8_wang2017epidemic} \cite{guo2016role} \\ \hline
Synthetic& 58 (92\%) & \cite{funk2009spread}\cite{JJA_jo2006immunization}\cite{weva13} \cite{wei2012competing} \cite{scata2018quantifying} \cite{guo2016role}\cite{JJ8_wang2017epidemic} \cite{sahneh2014competitive} \cite{joto17} \cite{dabo17} \cite{wano16} \cite{zhxu18} \cite{wuxi18} \cite{marceau2011modeling}\cite{sanz2014dynamics} \cite{sahneh2013may} \cite{azimi2016cooperative}\cite{JJ2_wang2016structural} \cite{JJ4_wei2016unified} \cite{JJ13_wei2018cooperative} \cite{JJ29_zhou2018immunizations} \cite{grgo14} \cite{jusa18} \cite{faji16} \cite{gule16} \cite{lich17} \cite{liwa16} \cite{kazh17} \cite{funk2010interacting} \cite{grgo13} \cite{czaplicka2016competition} \cite{wang2014asymmetrically} \cite{velasquez2017interacting} \cite{wang2016suppressing} \cite{scata2016impact} \cite{nicosia2017collective} \cite{nie2017impact} \cite{pan2018impact} \cite{pan2018impact2}  \cite{xia2019new} \cite{zhang2017epidemic} \cite{liu2016community} \cite{zhang2018new} \cite{yang2016impact} \cite{JJ3_gao2016competing} \cite{JJ15_gao2018dynamical} \cite{JJ20_huang2018global} \cite{JJ21_zang2018effects} \cite{JJ26_zhou2017numerical} \cite{JJ28_massaro2014epidemic} \cite{JJ30_chen2018optimal} \cite{JJ32_zheng2018interplay}\cite{JJ33_sagar2018effect} \cite{JJ35_guo2015two} \cite{JJE_eames2009networks} \cite{liti18} \cite{alla16} \cite{JJ34_chang2018co} \\ \hline

\end{tabular}
\end{center}
\caption{Type of network data used in the experiments}
\label{tab:NetworkData}
\end{table*}
As the experiments in the world of network science can be broadly divided into two main categories: (i) data-driven and (ii) simulation-based approaches, we expected more or less equal number of studies (i) where real-world networks where used and (ii) where models of networks where utilised to run simulation analysis. The very surprising finding that stroke us is that there is no reported research that looks into real-world multilayer networks with real multiple spreading processes propagating on them. This is especially interesting in the context of the newest publications as some appropriate datasets exist for a few years now \cite{jankowski2017multilayer}. 

Only few researchers use real-world multilayer networks for simulations of various spreads models (see \cite{wei2012competing}, \cite{weva13}, \cite{JJ8_wang2017epidemic}, \cite{guo2016role}, \cite{scata2018quantifying}), and although the spread used is still modelled using one of the traditional, not data-driven approaches, this setting is the closest to the real-world scenario.
In \cite{wei2012competing} and \cite{weva13} the authors used "\textit{Real-world enterprise composite network}", i.e. phone calls and text messages communication to create two layer network. Unfortunately, the dataset is not publicly available.
To validate simulations results Wang et. al. \cite{JJ8_wang2017epidemic} have used a Brightkite (location-based social networking service) dataset presented originally in \cite{cho2011friendship} and available for download on Stanford Large Network Dataset Collection\footnote{\url{https://snap.stanford.edu/data/}}. Based on this dataset, two layer network was built. The first layer (online communication network) was extracted based on the friendship in the Brightkite, and the second one (physical contact network) based on geographical proximity of two people (two people are connected in the physical contact network if they are within 200m from each other).
In \cite{guo2016role} the results of simulations are validated on the HIV1 network, which is built based on various genetic interactions for organisms. The dataset is available on CoMuNe Lab repository\footnote{\url{https://comunelab.fbk.eu/data.php}} and was originally presented in \cite{stark2006biogrid}. The original network has five layers but in \cite{guo2016role} only two of them were used (\textit{Physical association} and \textit{Direct interaction}).
Finally, authors of \cite{scata2018quantifying}, to validate the simulations results, have used the part of "\textit{machine classification dataset for suicide-related communication}" presented in \cite{burnap2015machine}. This dataset is not publicly available.

The vast majority of analysed papers performs experiments on artificially generated networks using existing models. For detailed list please see Table ~\ref{tab:NetworkData}. Most of them, to generate network layers, use well-defined network models such as Erdos--Renyi (ER) or Barabasi--Albert (BA). However, some of the studies use very unconventional methods to create network layers. Structures created in this way do not fit any model, their properties are unknown, and no explanation in regards to how they fit the reality is given. Most profound example, presented e.g. in \cite{grgo13} \cite{grgo14} \cite{zhang2017epidemic} \cite{JJ20_huang2018global} \cite{liti18}, is where authors generate one layer according to a selected existing model and then create a second layer by randomly adding 400 or 800 new links. Besides the fact that the new layer has unknown properties, this situation is hardly a reflection of a real-world scenario. The only reason to use such approach seems to be that authors want to create as big overlap between layers' edges as possible. Nevertheless, it is neither clear nor justified why such network generation process was employed. 

Among the models used for the two layer network generation the most popular combinations are Scale Free -- Scale Free networks 33\% of papers), ER -- ER networks 24\% and Scale Free -- ER networks (17\%) (see Tables \ref{tab:NetworkModels} and \ref{tab:NetworkModelsRef} for details). 

This lack or very limited explanation of why certain networks are used poses a very important questions in the context of the future research -- (i) how can we systematically explore different structures in the context of multiple spreads and (ii) how to ensure that the networks we used are representative in the context of specific research questions and at the same time can be generalized to be used to benchmark different approaches. These and others challenges are further explored in Section~\ref{sec:RoadMapGuidelines}.

\begin{table*}
\begin{center}
\begin{tabular}{l|rrrrrr}
\hline

 & Poisson (ER) & Exponetial & Small World & Scale Free & Regular & Other* \\ \hline
Poisson (ER) & \textbf{23.81}\% & \textbf{1.19}\% & 0.00\% & \textbf{16.67}\% & \textbf{1.19}\% & 0.00\% \\
Exponetial & & \textbf{1.19}\% & 0.00\% & \textbf{1.19}\% & \textbf{1.19}\% & 0.00\% \\
Small World & & & \textbf{5.95}\% &\textbf{ 1.19}\% & 0.00\% & 0.00\% \\
Scale Free & & & & \textbf{33.33}\% & \textbf{1.19}\% & \textbf{1.19}\% \\
Regular & & & & & \textbf{7.14}\% & 0.00\% \\
Other* & & & & & & \textbf{3.57}\% \\ \hline
\multicolumn{7}{r}{* For details please see Table } \ref{tab:NetworkModelsRef} \\
\end{tabular}
\end{center}
\caption{Network types combinations for two layer networks used} 
\label{tab:NetworkModels}
\end{table*}

\begin{table*}
\begin{center}
\begin{tabular}{l|l|p{13cm}}
\hline
\multicolumn{2}{c|}{Network models} & References \\ \hline

Poisson (ER) & Poisson (ER) & 
\cite{guo2016role} \cite{joto17} \cite{dabo17} \cite{wano16}\cite{zhxu18} \cite{wuxi18} \cite{marceau2011modeling} \cite{sanz2014dynamics} \cite{azimi2016cooperative} \cite{JJ29_zhou2018immunizations}\cite{jusa18} \cite{funk2010interacting} \cite{czaplicka2016competition} \cite{wang2016suppressing} \cite{yang2016impact}\cite{JJ21_zang2018effects}\cite{JJ26_zhou2017numerical}\cite{JJ28_massaro2014epidemic} \cite{JJ35_guo2015two}\cite{JJE_eames2009networks}\\ \hline
Poisson (ER) & Exponetial & \cite{jusa18} \\ \hline
Poisson (ER) & Scale Free & 
\cite{weva13} \cite{wei2012competing} \cite{guo2016role}\cite{JJ8_wang2017epidemic} \cite{sahneh2014competitive} \cite{marceau2011modeling}\cite{sahneh2013may}\cite{JJ2_wang2016structural}\cite{jusa18} \cite{liwa16} \cite{wang2014asymmetrically} \cite{nicosia2017collective} \cite{xia2019new}  \cite{JJ15_gao2018dynamical}   \\ \hline
Poisson (ER) & Regular & \cite{jusa18} \\ \hline
Exponetial & Exponetial & \cite{jusa18} \\ \hline
Exponetial & Scale Free & \cite{jusa18} \\ \hline
Exponetial & Regular & \cite{jusa18} \\ \hline
Small World & Small World & \cite{zhxu18}\cite{JJ13_wei2018cooperative} \cite{JJ29_zhou2018immunizations} \cite{pan2018impact}  \cite{liti18} \\ \hline
Small World & Scale Free & \cite{JJA_jo2006immunization} \\ \hline
Scale Free & Scale Free &  \cite{guo2016role} \cite{zhxu18}  \cite{sanz2014dynamics} \cite{JJ4_wei2016unified} \cite{JJ29_zhou2018immunizations} \cite{grgo14} \cite{jusa18} \cite{faji16} \cite{lich17} \cite{kazh17}    \cite{grgo13}\cite{scata2016impact} \cite{nie2017impact} \cite{pan2018impact} \cite{pan2018impact2}\cite{xia2019new} \cite{zhang2017epidemic} \cite{liu2016community} \cite{zhang2018new} \cite{JJ20_huang2018global} \cite{JJ21_zang2018effects} \cite{JJ26_zhou2017numerical} \cite{JJ28_massaro2014epidemic} \cite{JJ30_chen2018optimal} \cite{JJ32_zheng2018interplay}  \cite{JJ33_sagar2018effect} \cite{JJ35_guo2015two} \cite{liti18}\\ \hline
Scale Free & Regular & \cite{jusa18}\\ \hline
Scale Free & Other & \cite{gule16} - the information layer is generated using activity driven model. \\ \hline
Regular & Regular &
\cite{funk2009spread} \cite{jusa18} \cite{JJ26_zhou2017numerical}\cite{JJ28_massaro2014epidemic}   \cite{alla16}\cite{JJ34_chang2018co} \\ \hline 

Other & Other & \cite{weva13}\cite{wei2012competing} -- authors generated synthetic social networks using the Forest Fire, Random Walk, and Nearest Neighbor graph generation models proposed in \cite{sala2010measurement} to resemble real-world networks. \cite{velasquez2017interacting} -- firstly authors generate random links correlated between for both layers at the same time and secondly they randomly create uncorrelated links on each layer separately. In all cases there is no information on the resulting network topology. \\ \hline
\end{tabular}
\end{center}
\caption{Network types combinations for two layer networks used in each paper.}
\label{tab:NetworkModelsRef}
\end{table*}

\subsubsection{Full multilayer or just multiplex}
\label{subsec:multilayerVsMultiplex}
\begin{table*}
\begin{center}
\begin{tabular}{l|rp{13cm}}
\hline 

Network type & No. of papers & Reference \\ \hline
Multiplex & 53 (91\%) & \cite{funk2009spread}\cite{JJA_jo2006immunization} \cite{weva13} \cite{wei2012competing} \cite{scata2018quantifying} \cite{guo2016role} \cite{JJ8_wang2017epidemic} \cite{sahneh2014competitive}\cite{joto17} \cite{dabo17} \cite{zhxu18} \cite{wuxi18} \cite{marceau2011modeling}\cite{sanz2014dynamics} \cite{sahneh2013may} \cite{azimi2016cooperative} \cite{JJ2_wang2016structural}  \cite{JJ4_wei2016unified}\cite{JJ29_zhou2018immunizations} \cite{grgo14} \cite{jusa18} \cite{faji16} \cite{gule16} \cite{lich17}\cite{liwa16}\cite{kazh17}\cite{funk2010interacting}  \cite{grgo13} \cite{wang2014asymmetrically} \cite{velasquez2017interacting} \cite{wang2016suppressing} \cite{scata2016impact} \cite{nicosia2017collective}\cite{nie2017impact} \cite{pan2018impact} \cite{pan2018impact2} \cite{xia2019new} \cite{zhang2017epidemic} \cite{liu2016community} \cite{yang2016impact}\cite{JJ3_gao2016competing} \cite{JJ15_gao2018dynamical} \cite{JJ20_huang2018global} \cite{JJ21_zang2018effects} \cite{JJ26_zhou2017numerical} \cite{JJ28_massaro2014epidemic}\cite{JJ30_chen2018optimal} \cite{JJ32_zheng2018interplay}\cite{JJ33_sagar2018effect}  \cite{JJ35_guo2015two}\cite{JJE_eames2009networks}  \cite{liti18} \cite{JJ34_chang2018co} \\ \hline
Multilayer & 5 (9\%) & \cite{wano16}  \cite{JJ13_wei2018cooperative} \cite{czaplicka2016competition} \cite{zhang2018new}\cite{alla16} \\ \hline

\end{tabular}
\end{center}
\caption{Type of network used in the experiments}
\label{tab:MultilayerOrMultiplexNetwork}
\end{table*}

Multilayer networks are those where both nodes and edges can vary between the layers. Multiplex structures are a special instance of multilayer networks where only edges between layers can vary and the set of nodes remains the same for each layer \cite{kivela2014multilayer}, \cite{dickison2016multilayer}. The former ones are better reflection of real-life social networks, whereas the latter ones are useful representation used to limit the number of degrees of freedom when modelling complex networks and spread over them.

Out of all analysed papers only five (9\%) of them is using full multilayer networks \cite{wano16} \cite{alla16} \cite{JJ13_wei2018cooperative} \cite{czaplicka2016competition} \cite{zhang2018new}. This shows than the vast majority of the studies considers less complex case -- multiplex networks, or to be more specific node-aligned multiplex networks \cite{kivela2014multilayer} \cite{brodka2018quantifying}. In Table~\ref{tab:MultilayerOrMultiplexNetwork} we present which studies used which network type when modelling the multispread.

In reality, only a few networks are full multiplexes, and as multiplex networks are a simplification of multilayer case, they are not representative of a real-world scenario. For example, when one analyses the character of interaction between awareness and disease, one must consider that some people in the human contact network might not be present on the information network (e.g. Facebook). Additionally, some nodes which might be essential for spreading the information on the information network might not be present on contact network because for example they live in different geographical location or are bots forwarding the news and messages. All in all, modelling the system as multiplex network is a big assumption that should be dropped in future research.

\subsubsection{Number of Layers}
\label{subsec:numOfLayers}
To thoroughly investigate the composition of the structures used in the reviewed literature, we have decided to perform an analysis of the gathered information from two distinct perspectives. The first one focused on checking for how many layers the theoretical model of interactive spreading was proposed. The second one was to investigate on how many layers the model was tested during the experimental validation.

The vast majority (90\%) of introduced models were designed to work only on simple two layers networks. The only model designed for three layers networks was introduced in \cite{scata2018quantifying}. 

There are quite few papers \cite{funk2010interacting} \cite{sahneh2014competitive} \cite{JJ2_wang2016structural} \cite{sanz2014dynamics} \cite{nicosia2017collective}, which are introducing a general theoretical frameworks which are able to work on network with any number of layers. However, for experimental validation only three papers \cite{joto17} \cite{JJ2_wang2016structural} \cite{scata2018quantifying} are using three layers networks while the rest of them is limited to two layers networks (for details see Table~\ref{tab:NumberOfLayers}). 

Extending the number of layers in the experiments builds the complexity but at the same time is a must--have to fully understand the mechanisms behind multispread over multilayer networks.

\begin{table*}
\begin{center}
\begin{tabular}{p{5.5cm}p{1cm}p{1cm}|p{5.5cm}|p{1cm}}
\hline 
\multicolumn{5}{c}{Number of layers} \\ \hline
\multicolumn{3}{c|}{Theory} & \multicolumn{2}{c}{Experiments} \\ \hline
\multicolumn{1}{c}{2} & \multicolumn{1}{c}{3} & \multicolumn{1}{c|}{n} & \multicolumn{1}{c}{2} & \multicolumn{1}{c}{3} \\ \hline
53 (88\%) & 1 (2\%) & 6 (10\%) & 56 (95\%) & 3 (5\%) \\
\cite{funk2009spread}\cite{JJA_jo2006immunization} \cite{weva13} \cite{wei2012competing}\cite{JJ8_wang2017epidemic}  \cite{guo2016role}\cite{sahneh2014competitive} \cite{dabo17}  \cite{wano16} \cite{zhxu18}\cite{wuxi18}  \cite{marceau2011modeling} \cite{sahneh2013may}  \cite{azimi2016cooperative}\cite{JJ4_wei2016unified}\cite{JJ13_wei2018cooperative} \cite{JJ29_zhou2018immunizations} \cite{grgo14} \cite{jusa18} \cite{faji16} \cite{gule16} \cite{lich17}\cite{liwa16} \cite{kazh17} \cite{funk2010interacting} \cite{grgo13} \cite{czaplicka2016competition} \cite{wang2014asymmetrically} \cite{velasquez2017interacting} \cite{wang2016suppressing} \cite{scata2016impact} \cite{nie2017impact} \cite{pan2018impact} \cite{pan2018impact2} \cite{xia2019new} \cite{zhang2017epidemic} \cite{liu2016community} \cite{zhang2018new} \cite{yang2016impact}\cite{JJ3_gao2016competing}\cite{JJ15_gao2018dynamical} \cite{JJ20_huang2018global} \cite{JJ21_zang2018effects} \cite{JJ26_zhou2017numerical} \cite{JJ28_massaro2014epidemic} \cite{JJ30_chen2018optimal} \cite{JJ32_zheng2018interplay} \cite{JJ33_sagar2018effect} \cite{JJ35_guo2015two}\cite{JJE_eames2009networks} \cite{liti18} \cite{alla16}\cite{JJ34_chang2018co}
& \cite{scata2018quantifying}
& \cite{sahneh2014competitive} \cite{joto17} \cite{sanz2014dynamics}\cite{JJ2_wang2016structural}  \cite{funk2010interacting} \cite{nicosia2017collective} 
& \cite{funk2009spread}\cite{JJA_jo2006immunization} \cite{weva13} \cite{wei2012competing}\cite{JJ8_wang2017epidemic}  \cite{guo2016role}\cite{sahneh2014competitive} \cite{joto17} \cite{dabo17}  \cite{wano16} \cite{zhxu18}\cite{wuxi18}  \cite{marceau2011modeling}\cite{sanz2014dynamics} \cite{sahneh2013may}  \cite{azimi2016cooperative}\cite{JJ4_wei2016unified}\cite{JJ13_wei2018cooperative} \cite{JJ29_zhou2018immunizations} \cite{grgo14} \cite{jusa18} \cite{faji16} \cite{gule16} \cite{lich17}\cite{liwa16} \cite{kazh17} \cite{funk2010interacting} \cite{grgo13} \cite{czaplicka2016competition} \cite{wang2014asymmetrically} \cite{velasquez2017interacting} \cite{wang2016suppressing} \cite{scata2016impact} \cite{nicosia2017collective} \cite{nie2017impact} \cite{pan2018impact} \cite{pan2018impact2} \cite{xia2019new} \cite{zhang2017epidemic} \cite{liu2016community} \cite{zhang2018new} \cite{yang2016impact}\cite{JJ3_gao2016competing}\cite{JJ15_gao2018dynamical} \cite{JJ20_huang2018global} \cite{JJ21_zang2018effects} \cite{JJ26_zhou2017numerical} \cite{JJ28_massaro2014epidemic} \cite{JJ30_chen2018optimal} \cite{JJ32_zheng2018interplay} \cite{JJ33_sagar2018effect} \cite{JJ35_guo2015two}\cite{JJE_eames2009networks} \cite{liti18} \cite{alla16}\cite{JJ34_chang2018co}
& \cite{scata2018quantifying} \cite{joto17} \cite{JJ2_wang2016structural}  \\ \hline
\end{tabular}
\end{center}
\caption{The number of layers in the multilayer networks used in each paper}
\label{tab:NumberOfLayers}
\end{table*}

\subsubsection{Edges between layers}
\label{subsec:edgesBetweenLayers}
\begin{table*}
\begin{center}
\begin{tabular}{p{2cm}|p{13cm}}
\hline 
\multicolumn{2}{l}{Interlayer edges considered?} \\ \hline
Yes & No \\ \hline
3 (4\%) & 55 (95\%)\\
\cite{JJ13_wei2018cooperative} \cite{czaplicka2016competition} \cite{alla16}
& 
\cite{funk2009spread} \cite{JJA_jo2006immunization}\cite{weva13} \cite{wei2012competing}  \cite{scata2018quantifying} \cite{JJ8_wang2017epidemic} \cite{guo2016role} \cite{sahneh2014competitive}\cite{joto17}  \cite{dabo17}\cite{wano16}\cite{zhxu18} \cite{wuxi18}\cite{marceau2011modeling}\cite{sanz2014dynamics}\cite{sahneh2013may} \cite{azimi2016cooperative}\cite{JJ2_wang2016structural}  \cite{JJ4_wei2016unified} \cite{JJ13_wei2018cooperative}\cite{JJ29_zhou2018immunizations}\cite{grgo14} \cite{jusa18} \cite{faji16} \cite{gule16}\cite{lich17} \cite{liwa16}\cite{kazh17}\cite{funk2010interacting}\cite{grgo13}  \cite{wang2014asymmetrically} \cite{velasquez2017interacting} \cite{wang2016suppressing} \cite{scata2016impact} \cite{nicosia2017collective}\cite{nie2017impact} \cite{pan2018impact} \cite{pan2018impact2} \cite{xia2019new}  \cite{zhang2017epidemic} \cite{liu2016community} \cite{zhang2018new} \cite{yang2016impact}\cite{JJ3_gao2016competing} \cite{JJ15_gao2018dynamical} \cite{JJ20_huang2018global}\cite{JJ21_zang2018effects} \cite{JJ26_zhou2017numerical}\cite{JJ28_massaro2014epidemic}   \cite{JJ30_chen2018optimal} \cite{JJ32_zheng2018interplay}\cite{JJ35_guo2015two}\cite{JJE_eames2009networks} \cite{liti18} \cite{JJ34_chang2018co}
\\ \hline

\end{tabular}
\end{center}
\caption{Usage of the interlayer edges}
\label{tab:InterlayerEdges}
\end{table*}
Another element, which builds the complexity of the topic, is the existence of the edges between layers. 
In most cases, researchers do not consider additional interlayer edges, what most probably is a result of using multiplex networks where those edges are not needed since each node is present on all layers. However, some papers \cite{alla16} \cite{JJ13_wei2018cooperative} \cite{czaplicka2016competition}, propose models for, and perform experiments on, full multilayer networks with interlayer edges (see Table~\ref{tab:InterlayerEdges} for papers falling in respective categories in regards to the existence of interlayer edges).
In \cite{czaplicka2016competition}, authors propose to have $M$ interlayer edges between two layers that randomly connect nodes between two layers. Thus, the change in the node opinion is affected both by the neighbours in its layer and by the neighbours in the other layer. 
Wei \cite{JJ13_wei2018cooperative} connects two homogeneous networks with random interlayer links without degree correlations. Nodes in the first layer can be infected by connected infected neighbours from the second layer.
Similarly to Wei, authors in \cite{alla16} create interlayer edges connecting each vertex in one layer to one vertex which is selected randomly from the other layer. Those connections enable nodes from layer where the opinion is formed to influence the nodes on the second layer where the decision--making process is formed.

As in the case of network type (multiplex vs multilayer) used in the reviewed studies, also the existence of interlayer edges, researchers tend to go for the option that reduces the complexity of the problem. While this is an obvious and reasonable approach to start with, the natural next and very much needed step is to investigate also more complex settings.

\subsubsection{External influence}
\label{subsec:ExternalInfluence}
\begin{table*}
\begin{center}
\begin{tabular}{p{2cm}|p{13cm}}
\hline 
\multicolumn{2}{l}{Is external environment considered?} \\ \hline
Yes & No \\ \hline
7 (12\%) & 52 (88\%)\\
\cite{JJ29_zhou2018immunizations} \cite{grgo14} \cite{kazh17} \cite{xia2019new}\cite{guo2016role} \cite{JJ3_gao2016competing} \cite{JJ21_zang2018effects} \cite{JJE_eames2009networks} 
& 
\cite{funk2009spread}\cite{JJA_jo2006immunization}\cite{weva13} \cite{wei2012competing}\cite{scata2018quantifying} \cite{JJ8_wang2017epidemic} \cite{sahneh2014competitive} \cite{joto17} \cite{dabo17} \cite{wano16} \cite{zhxu18}\cite{wuxi18}\cite{marceau2011modeling}\cite{sanz2014dynamics} \cite{sahneh2013may}\cite{azimi2016cooperative} \cite{JJ2_wang2016structural} \cite{JJ4_wei2016unified} \cite{JJ13_wei2018cooperative}\cite{jusa18} \cite{faji16} \cite{gule16} \cite{lich17}\cite{liwa16}\cite{kazh17}\cite{funk2010interacting} \cite{grgo13}\cite{czaplicka2016competition}  \cite{wang2014asymmetrically} \cite{velasquez2017interacting} \cite{wang2016suppressing} \cite{scata2016impact} \cite{nicosia2017collective}\cite{nie2017impact} \cite{pan2018impact} \cite{pan2018impact2} \cite{zhang2017epidemic} \cite{liu2016community} \cite{zhang2018new} \cite{yang2016impact}\cite{JJ15_gao2018dynamical} \cite{JJ20_huang2018global} \cite{JJ26_zhou2017numerical} \cite{JJ28_massaro2014epidemic} \cite{JJ30_chen2018optimal} \cite{JJ32_zheng2018interplay} \cite{JJ33_sagar2018effect}  \cite{JJ35_guo2015two}\cite{liti18}\cite{alla16} \cite{JJ34_chang2018co}\cite{saumell2012epidemic} \\ \hline

\end{tabular}
\end{center}
\caption{The number of papers which considered the external environmente}
\label{tab:ExternalEnviroment}
\end{table*}
In any complex system, one of the biggest challenges is to understand and model the interaction with, and the influence of, the external environment. Multispread over multilayer networks is no different in this respect, as it is a classic example of a complex process propagating over a complex system. 

Most of the reviewed studies assume that the system where the spreading process takes place is isolated and there is no interaction between the system and the external environment (see Table~\ref{tab:ExternalEnviroment} for the comprehensive list of relevant papers)). This is inline with the observations from the previous sections -- researchers reduce the complexity of the multispread over multilayer networks problem, which is relatively new altogether. This is a natural inclination, as we first need to learn to walk before we run.

Having said that, some researchers \cite{grgo14} \cite{kazh17} \cite{JJ29_zhou2018immunizations} \cite{JJ3_gao2016competing} \cite{JJE_eames2009networks} \cite{xia2019new} enrich their models by taking into account such external factors as the influence of the media or global immunizations strategies. For example, in \cite{grgo14} and \cite{xia2019new} authors simulate the influence of media by creating, in the information layer, an artificial node which is connected to every other node in that layer. In regular time intervals, this node sends information about the disease to all nodes that belong to the information layer and because of that message, informed nodes, with certain low probability, can alter their state from being unaware to aware. A similar approach can be found in \cite{kazh17} where mass media influences the awareness level depending on how many people are infected. The individual probability of becoming self-aware increases if more people around is infected.

Zhou \cite{JJ29_zhou2018immunizations} simulates the external influence in the form of immunization of important nodes in the network. Another research shows that combined self--protection with external information is an effective strategy to decrease epidemic spreading \cite{JJ3_gao2016competing}. Government information campaigns focused on vaccination can be represented in simulations by activation of a fraction of nodes with initial opinion \cite{JJE_eames2009networks}.

External environment and its influence on the diffusion process plays a pivotal role in how different, multiple phenomena propagate over multilayer networks, and it cannot be neglected in the future research in this field.


\subsection{How it spreads?}
\label{sec:how}
The discussion about the system in which the spread takes place and how the external environment can influence the spread showed how complex the problem is. However, the complexity is not only built by the medium where the diffusion happens, but also by the processes and interactions between them. Thus, the next element that we investigate answers the question "how?". We focus on how individual spreading processes and the interaction between them are modelled. 

\subsubsection{Spread models}

\begin{table*}
\begin{center}
\begin{tabular}{p{10cm}|p{5cm}}\hline
\multicolumn{2}{c}{Is the same spreading model on all layers?}\\ \hline
Yes -- 39 papers (67\%)&No -- 19 papers (33\%)\\
\cite{weva13} \cite{wei2012competing}\cite{scata2018quantifying} \cite{guo2016role}\cite{JJ8_wang2017epidemic}  \cite{sahneh2014competitive}\cite{joto17} \cite{dabo17} \cite{wano16}\cite{zhxu18}\cite{marceau2011modeling}\cite{sanz2014dynamics} \cite{azimi2016cooperative} \cite{JJ2_wang2016structural}\cite{JJ4_wei2016unified}\cite{JJ13_wei2018cooperative}\cite{JJ29_zhou2018immunizations}\cite{grgo14} \cite{jusa18} \cite{faji16} \cite{gule16}\cite{kazh17}\cite{funk2010interacting}  \cite{grgo13}\cite{scata2016impact} \cite{nie2017impact} \cite{pan2018impact} \cite{zhang2017epidemic} \cite{zhang2018new}\cite{JJ3_gao2016competing} \cite{JJ15_gao2018dynamical} \cite{JJ20_huang2018global} \cite{JJ21_zang2018effects} \cite{JJ26_zhou2017numerical}\cite{JJ28_massaro2014epidemic}  \cite{JJ30_chen2018optimal} \cite{JJ32_zheng2018interplay} \cite{JJ33_sagar2018effect} \cite{JJ35_guo2015two} \cite{JJ34_chang2018co} 
&
\cite{funk2009spread} \cite{JJA_jo2006immunization}\cite{wuxi18}\cite{sahneh2013may} \cite{lich17}  \cite{liwa16}\cite{czaplicka2016competition} \cite{wang2014asymmetrically} \cite{velasquez2017interacting} \cite{wang2016suppressing} \cite{nicosia2017collective} \cite{pan2018impact2}\cite{xia2019new} \cite{liu2016community} \cite{yang2016impact}\cite{JJE_eames2009networks}\cite{liti18}\cite{alla16} \\ \hline

\end{tabular}
\end{center}
\caption{Papers which use the same model on all layers and those which use different model for each layer.}
\label{tab:thesamemodels}
\end{table*}

The first element to look at is what spreading models are used when more than one spread takes place in multilayer network. Not surprisingly, analysed papers mainly focus on epidemic models previously used in the context of a single virus spreading within single layer. Most of the papers consider simple epidemic models like SIR (Susceptible-Infectious-Recovered) and SIS (Susceptible-Infectious-Susceptible) and their variations with the same model used on all layers (see Table~\ref{tab:spreadmodels} for details). Note that, \textbf{41\%} of them consider the SIS model on both layers. In case of papers related to epidemics and awareness, the layer with awareness adopts SIS but is interpreted as Unaware-Aware-Unaware (UAU) model. Since it maintains the core characteristics of the SIS model, we treat it as the same class of models. The SIR model is used in \textbf{16\%} of papers in both layers. Similarly, to SIS, SIR also has been adopted for awareness propagation and named Unaware-Aware-Faded (UAF) which we treat as SIR model for the same reason as we treat UAU as SIS model. 

\textbf{9\%} of papers use on both layers extension of SIS model towards multiple contagions in a form of SI1SI2S model and \textbf{7\%} of papers use Threshold models (TM). Only single papers use other models, such as Independent Cascade Model (ICM) \cite{liu2016community}, opinion formation model \cite{JJE_eames2009networks}, Random Walk \cite{nicosia2017collective}, Kuramoto model \cite{nicosia2017collective}, Voter model \cite{velasquez2017interacting} and Contact Process (asynchronous SIS model) \cite{velasquez2017interacting}. Apart from presented approaches other models like GACS \cite{JJ21_zang2018effects}, LACS \cite{JJ35_guo2015two} and M-model \cite{alla16} are used in \textbf{5\%} of papers.

One-third of papers use different spreading models on each layer (Table \ref{tab:thesamemodels}). A combination of SIS and SIR model was used for two layers spreading in \textbf{5\%} of works. SIS model was used together with threshold models in \textbf{5\%} of papers including \cite{czaplicka2016competition} \cite{pan2018impact2} \cite{guo2016role} and in \textbf{2\%} of works together with SIRV model \cite{wang2016suppressing}.

 Combination of SIR with SIRV was applied in \textbf{5\%} of papers \cite{lich17} \cite{liwa16} \cite{wang2014asymmetrically}. 
SIR was used together with ICM model in \cite{liu2016community} and with opinion spreading model \cite{JJE_eames2009networks}. Other works combined random walk on first layer with Kuramoto model on the second layer \cite{nicosia2017collective} and voter model with contact process \cite{velasquez2017interacting}. 

The analysis shows that there is a clear tendency that is very similar to what we can see in the "where" section (\ref{sec:where}). Researchers tend to simplify the problem and investigate the interaction between processes using, in the majority of cases, simple, epidemic models that are well understood in the one layer scenario. While this is the right thing to do, there is also a need to depart from those models and look more into data-driven models that can adapt over time. 

\begin{table*}
\begin{center}
\begin{tabular}{l|ccccc}
\hline
&SIS (UAU) & SIR (UAF) & SI1SI2S & TM & Other* \\ \hline
SIS (UAU)&\textbf{41.38}\%&\textbf{5.17}\%\% & 0.00 & \textbf{5.17}\% & 0.00 \\
SIR (UAF) & 0.00 &\textbf{15.52}\% & 0.00 & 0.00 & \textbf{5.17}\% \\
SIRV & \textbf{1.72}\% & \textbf{5.17}\%\ & 0.00& 0.00& 0.00 \\
SI1SI2S & 0.00& 0.00& \textbf{8.62}\%& 0.00 & 0.00 \\
TM & 0.00& 0.00& 0.00 & \textbf{1.72}\% & 0.00 \\
Other* & 0.00& 0.00& 0.00 & 0.00 & \textbf{10.34}\% \\ \hline
\multicolumn{6}{r}{* - for details please see Table \ref{tab:model}}\\
\end{tabular}
\end{center}
\caption{Spreading models used in the papers.}
\label{tab:spreadmodels}
\end{table*}

\begin{table*}
\begin{center}
\begin{tabular}{l|l|p{11cm}}
\hline
\multicolumn{2}{c|}{Spreading models} & References \\ \hline
SIS & SIS & 
\cite{JJ8_wang2017epidemic}\cite{joto17} \cite{dabo17}\cite{wuxi18}  \cite{JJ4_wei2016unified}  \cite{JJ13_wei2018cooperative} \cite{sanz2014dynamics}\cite{grgo14}  \cite{jusa18} \cite{faji16} \cite{gule16}  \cite{kazh17} \cite{grgo13} \cite{nie2017impact} \cite{pan2018impact} \cite{zhang2017epidemic} \cite{zhang2018new}\cite{JJ15_gao2018dynamical} \cite{JJ20_huang2018global} \cite{JJ26_zhou2017numerical} \cite{JJ28_massaro2014epidemic} \cite{JJ30_chen2018optimal} \cite{JJ33_sagar2018effect}\cite{liti18}\\ \hline
SIS & SIR & 
\cite{xia2019new} \cite{yang2016impact}
\cite{JJ32_zheng2018interplay} \\ \hline
SIS & TM & 
\cite{guo2016role} \cite{czaplicka2016competition} \cite{pan2018impact2} \\ \hline
SIS & SIRV & 
\cite{wang2016suppressing} \\ \hline
SIR & SIR & 
\cite{JJA_jo2006immunization}\cite{zhxu18} \cite{marceau2011modeling} \cite{sanz2014dynamics} \cite{azimi2016cooperative}\cite{JJ29_zhou2018immunizations}\cite{funk2010interacting}  \cite{scata2016impact}\cite{JJ3_gao2016competing} \\ \hline
SIR & SIRV & 
\cite{lich17} \cite{liwa16}
\cite{wang2014asymmetrically} \\ \hline
SIR & ICM & 
\cite{liu2016community} \\ \hline
SIR & Opinion & 
\cite{JJE_eames2009networks} \\ \hline
SIR & Awareness model & 
\cite{funk2009spread} \\ \hline
SI1SI2S & SI1SI2S & 
\cite{weva13} \cite{wei2012competing} \cite{sahneh2014competitive} \cite{wano16}
\cite{sahneh2013may} \\ \hline
Random Walk & Kuramoto & 
\cite{nicosia2017collective} \\ \hline
Voter Model & Contact & 
\cite{velasquez2017interacting} \\ \hline
SI1S2 & SI1S2 & 
\cite{JJ2_wang2016structural} \\ \hline
TM & TM & 
\cite{JJ34_chang2018co} \\ \hline
M-model & M-model & \cite{alla16} \\ \hline
LACS & LACS & \cite{JJ35_guo2015two} \\ \hline
GACS & GACS & \cite{JJ21_zang2018effects} \\ \hline
\end{tabular}
\end{center}
\caption{Spreading models combinations used on two layer networks.}
\label{tab:model}
\end{table*}

\subsubsection{Spread switches layers}
\begin{table*}
\begin{center}
\begin{tabular}{p{2cm}|p{13cm}}
\hline
\multicolumn{2}{l}{Are the processes able to switch layer?}\\ \hline
Yes - 5 (9\%) & No - 53 (91\%) \\
\cite{JJA_jo2006immunization}\cite{weva13} \cite{wei2012competing} \cite{JJ2_wang2016structural} \cite{JJ13_wei2018cooperative} 
&
 \cite{funk2009spread} \cite{scata2018quantifying}\cite{guo2016role}\cite{JJ8_wang2017epidemic} \cite{sahneh2014competitive}\cite{joto17} \cite{dabo17} \cite{wano16} \cite{zhxu18}\cite{wuxi18}\cite{marceau2011modeling}\cite{sanz2014dynamics} \cite{sahneh2013may} \cite{azimi2016cooperative} \cite{JJ4_wei2016unified} \cite{JJ29_zhou2018immunizations}\cite{grgo14} \cite{jusa18} \cite{faji16} \cite{gule16} \cite{lich17} \cite{liwa16}\cite{kazh17}\cite{funk2010interacting} \cite{grgo13} \cite{czaplicka2016competition}\cite{wang2014asymmetrically} \cite{velasquez2017interacting} \cite{wang2016suppressing}\cite{scata2016impact}  \cite{nicosia2017collective} \cite{nie2017impact} \cite{pan2018impact} \cite{pan2018impact2}\cite{xia2019new} \cite{zhang2017epidemic} \cite{liu2016community} \cite{zhang2018new} \cite{yang2016impact} \cite{JJ3_gao2016competing} \cite{JJ15_gao2018dynamical} \cite{JJ20_huang2018global} \cite{JJ21_zang2018effects} \cite{JJ26_zhou2017numerical}\cite{JJ28_massaro2014epidemic}\cite{JJ30_chen2018optimal} \cite{JJ32_zheng2018interplay} \cite{JJ33_sagar2018effect}\cite{JJ35_guo2015two}\cite{JJE_eames2009networks} \cite{liti18} \cite{alla16} \cite{JJ34_chang2018co} \\ 
 \hline
\end{tabular}
\end{center}
\caption{The papers in which the process is and is not able to switch between layers.}
\label{tab:layerSwitch}
\end{table*}

An important element when considering multiple spreads in the multilayer networked environment is the ability for spreads to switch layers. It is intriguing, especially in the context of information/gossip and alike phenomena spreads, as it shows the natural way in which the diffusion in social systems happens, e.g. some things propagate over one system (e.g. Twitter) and all of a sudden they jump into another (e.g. Facebook).

Having said that, there is only limited analysis were one spread can move from one layer to another (only \textbf{9\%} of papers, see Table~\ref{tab:layerSwitch}. Please note, that we differentiate switching from coupling. We understand switching layers as the ability of moving the contagion ("what?" from Section~\ref{sec:what}) from one layer to another. Coupling is not equal to spread switching layers - coupling effectively means that being subject to process (e.g. infected) on one layer can cause that for this node the second process is triggered on the other layer (e.g. node infected on the contact layer becomes aware on the communication layer). In general, most of works show separate layers and different types of content is transmitted on each of them. For example, pathogen can be transmitted only within real contacts network not within information network based on electronic communication and social media. 

Switching layers is possible if the layers can transmit the same content, for example information. In \cite{weva13} and \cite{wei2012competing} authors conduct cross-contamination experiment where, with a certain probability, one meme can jump from one to the other layer (from phone calls layer to SMS layer or vice versa) and spreads there. Conceptually, similar approach is presented in \cite{JJ2_wang2016structural} and assumes that viral agent can be spread by the infected node to its neighbours in all layers. Analogous setup for two layers based on prevention and infection networks assumed infection of susceptible nodes by infected neighbours on any layer \cite{JJA_jo2006immunization}. Another study analysed the role of interlayer correlations and interconnections for scenarios when a single node in the susceptible state can be infected by neighbours on different layers simultaneously \cite{JJ13_wei2018cooperative}. Recovery was modeled in the same way, and infected node can recover on different layers with a certain probability. 

Outstanding \textbf{91\%} of works assumes that spreading of each "what" (see sec \ref{sec:what}) takes place only within one layer. The fact that only \textbf{9\%} pf research considers the possibility for the spread to switch layers, shows again the natural tendency to simplify the problem and break it to more manageable pieces. Yet again, the complexity that is brought into the equation by the phenomena that can switch the layers cannot be neglected. 

\subsubsection{Co-infection of nodes}
\begin{table*}
\begin{center}
\begin{tabular}{p{12cm}|p{3cm}}
\hline
\multicolumn{2}{c}{Is it possible that the node is affected by both processes at the same time?}\\ \hline
Yes -- 49 papers (80\%)& No -- 13 papers (20\%)\\
\cite{funk2009spread}\cite{scata2018quantifying}\cite{guo2016role} \cite{JJ8_wang2017epidemic}\cite{joto17} \cite{dabo17}\cite{zhxu18} \cite{wuxi18} \cite{marceau2011modeling} \cite{sanz2014dynamics} \cite{azimi2016cooperative}\cite{JJ4_wei2016unified} \cite{JJ13_wei2018cooperative}  \cite{JJ29_zhou2018immunizations} \cite{grgo14}  \cite{jusa18} \cite{faji16} \cite{gule16} \cite{lich17} \cite{liwa16} \cite{kazh17}  \cite{funk2010interacting} \cite{grgo13}\cite{wang2014asymmetrically} \cite{velasquez2017interacting} \cite{wang2016suppressing} \cite{scata2016impact} \cite{nicosia2017collective}  \cite{nie2017impact} \cite{pan2018impact} \cite{pan2018impact2} \cite{xia2019new} \cite{zhang2017epidemic} \cite{liu2016community} \cite{zhang2018new} \cite{yang2016impact} \cite{JJ3_gao2016competing}  \cite{JJ15_gao2018dynamical} \cite{JJ20_huang2018global} \cite{JJ21_zang2018effects} \cite{JJ26_zhou2017numerical}\cite{JJ28_massaro2014epidemic}\cite{JJ30_chen2018optimal} \cite{JJ32_zheng2018interplay} \cite{JJ33_sagar2018effect} \cite{JJ35_guo2015two} \cite{liti18} \cite{alla16}\cite{JJ34_chang2018co} 
&
\cite{JJA_jo2006immunization} \cite{weva13} \cite{wei2012competing} \cite{sahneh2014competitive} \cite{joto17} \cite{dabo17} \cite{wano16}\cite{marceau2011modeling} \cite{sahneh2013may}\cite{JJ2_wang2016structural}\cite{funk2010interacting} \cite{czaplicka2016competition}\cite{JJE_eames2009networks}  
\\ \hline
\end{tabular}
\end{center}
\caption{The papers in which the node can and cannot be affected by more than one process at the time.}
\label{tab:coinfection}
\end{table*}

Next element we analyse is the ability of nodes to be in "activated" state at the same time by several processes. It can be treated as inclusive adoption \cite{JJ34_chang2018co} which means that a node can adopt many, different things/phenomena. For example, if information is spreading within the network, a single node can possess different, sometimes contradicting, information at the same time. Similar situation happens when one node can be co-infected by several contagions. In most of the cases (\textbf{80\%} of papers) authors assumed that co-infection is possible. 

The most common scenario, when co-infection on nodes is possible, is when epidemic spreads on one layer and awareness on another. Then a node can be both aware of and infected at the same time. The element that should be emphasized is that in some cases there are three possible states AI (aware-infected), AS (aware-susceptible), and US (unaware-susceptible) \cite{faji16} \cite{gule16} and sometimes there are four states with additional UI (unaware-infected) state \cite{kazh17}. In the latter case, the fact that a node is infected does not necessarily imply that it is aware at the same time. 

There is also an option when co-infection can happen because the protective process does not give the full immunity and a given node can be still infected \cite{funk2010interacting} \cite{marceau2011modeling}.

In cases when we have two epidemics spreading at the same time, usually this means that a given node can be infected only by one epidemic \cite{funk2010interacting} \cite{sahneh2014competitive} \cite{saumell2012epidemic} \cite{marceau2011modeling}. Similar case is when the memes spread - in \cite{weva13} and \cite{wei2012competing} one node can posses single meme at a time. 

In papers where co-infection was not possible authors assume that viral agent kills any other and node can be activated by single content at the same time only \cite{JJ2_wang2016structural}, nodes can be either in infected or in immune state \cite{JJA_jo2006immunization}, one physical layer is based on children (diseases spread) and second is based on parents and information network for awareness spread) \cite{JJE_eames2009networks}. Also, in the case where each of the network layers is modelled as a separate transmission channel of the contagions, each node will be infected only by one spread at a given time \cite{sahneh2014competitive}. Others scenarios are analysed in \cite{marceau2011modeling} and \cite{czaplicka2016competition} where each node can be infected by only one spread at a time, but both spreads are the same.

Apart from focus on a single approach, studies presented in \cite{joto17} \cite{dabo17} showed possible infection of a single node by multiple contagions and compared it with competitive version where co-infection was not possible. 

\subsubsection{Interaction between processes}
\begin{table*}
\begin{center}
\begin{tabular}{l|rp{12cm}} \hline
Interaction type& No. of papers & References \\ \hline
Supporting & 7 (11\%)& \cite{dabo17} \cite{zhxu18} \cite{sanz2014dynamics} \cite{azimi2016cooperative} \cite{JJ13_wei2018cooperative}  \cite{nicosia2017collective}\cite{alla16} \\

Competing & 24 (36\%) & \cite{weva13} \cite{wei2012competing} \cite{sahneh2014competitive}\cite{joto17} \cite{dabo17} \cite{wano16}\cite{zhxu18}\cite{marceau2011modeling}\cite{sanz2014dynamics} \cite{sahneh2013may} \cite{grgo14}\cite{jusa18} \cite{funk2010interacting}\cite{wang2014asymmetrically} \cite{velasquez2017interacting} \cite{scata2016impact} \cite{nicosia2017collective} \cite{zhang2017epidemic}\cite{JJ21_zang2018effects}\cite{JJ30_chen2018optimal} \cite{JJE_eames2009networks}\cite{liti18}\cite{alla16}  
\\
Mixed& 30 (46\%) & 
\cite{funk2009spread}\cite{JJA_jo2006immunization} \cite{scata2018quantifying}\cite{guo2016role}  \cite{JJ8_wang2017epidemic}\cite{JJ2_wang2016structural}\cite{JJ4_wei2016unified}\cite{JJ29_zhou2018immunizations}  \cite{faji16} \cite{gule16} \cite{lich17} \cite{liwa16} \cite{kazh17}\cite{grgo13} \cite{wang2016suppressing} \cite{nie2017impact} \cite{pan2018impact} \cite{pan2018impact2} \cite{xia2019new}\cite{liu2016community}\cite{zhang2018new} \cite{yang2016impact}\cite{JJ3_gao2016competing}  \cite{JJ15_gao2018dynamical} \cite{JJ20_huang2018global} \cite{JJ26_zhou2017numerical}\cite{JJ28_massaro2014epidemic}
\cite{JJ32_zheng2018interplay} \cite{JJ33_sagar2018effect} \cite{JJ35_guo2015two} 
\\
None & 5 (7\%) & \cite{joto17} \cite{dabo17} \cite{wuxi18}\cite{czaplicka2016competition} \cite{JJ34_chang2018co}  \\ \hline
\end{tabular}
\end{center}
\caption{Types of interaction between spreading processes.}
\label{tab:interaction}
\end{table*}

Interactions between multiple contagions within networks can mainly take a form of competition or supporting actions. In the area of epidemiology, one disease can enhance or inhibit spread of another one \cite{zhxu18}. It can be observed in the area of viral marketing and competing products, interactions between awareness and infectious disease or suppression of disease spread by the immunization process. Interactions can be based on interdependency or cooperation and competition or antagonism \cite{dabo17}. Processes can also be interdependent and competitive simultaneously. For example, new similar hi-tech products create demand for new services which shows interdependence, but competition among them also takes place. 

Another possibility is to consider inclusive and exclusive adoption with the ability to possess multiple information or viruses by a single node at the same time or not respectively \cite{JJ34_chang2018co}. Inclusive adoption can be observed at the market when a consumer must adopt the first product prior to adopting the second one. For exclusive adoption, the first product will be replaced by the second one. 

Several papers conduct experiments for both: (i) supporting and (ii) competing scenarios. A general framework for interacting processes on multilayer networks was proposed in \cite{nicosia2017collective}. It enables to define what kind of interaction there is between both processes. A similar approach, where depending on the setting different types of interactions between processes can be defined, is propose in \cite{dabo17}. 

Apart from those few overarching studies, most of the research focuses on experiments falling into one of the following categories: (i)~supporting, (ii)~competing or (iii)~mixed approaches.

\textbf{(i) Supporting/Collaborating/Cooperative approaches.} The review shows that \textbf{11\%} of the papers investigates supporting processes. Cooperative spreading processes are presented by \cite{JJ13_wei2018cooperative} with focus on role of layers structures and their correlations. Interplay between processes is observed for opinion formation and decision making, where the opinion of public about certain issue is taken into account during the decision making process that takes place at the higher level \cite{alla16}. Epidemics on multiplex networks can take cooperative form \cite{azimi2016cooperative}. As a result dynamics and coverage of one disease can be increased by other diseases spreading on the same network. One disease can be a consequence of being infected with another one \cite{sanz2014dynamics}. For example the number of people with tuberculosis is much higher among people with HIV \cite{azimi2016cooperative}

\textbf{(ii) Competing/Suppressing approaches.} In \textbf{36\%} of the papers competing processes are modelled and analysed. Competition between processes was analysed for memes \cite{weva13} \cite{wei2012competing} and extended towards generalised models for other content \cite{sahneh2013may}. Protective spread like cure or immunity can compete with virus or diseases and this scenario is explored in \cite{funk2010interacting} and \cite{marceau2011modeling}. Also, two competing viruses \cite{sahneh2014competitive} \cite{zhxu18} \cite{joto17} and coupling between both diffusion processes \cite{sanz2014dynamics} were analysed. 

Another studies analyse competition of epidemics and awareness \cite{grgo14} \cite{jusa18} \cite{JJ32_zheng2018interplay} \cite{JJ33_sagar2018effect} \cite{nie2017impact} \cite{pan2018impact} \cite{pan2018impact2} \cite{zhang2017epidemic} \cite{wuxi18}. Apart from generalised models they focus on awareness cascades \cite{JJ35_guo2015two} and global awareness \cite{JJ21_zang2018effects}. Competition takes place also for information diffusion to prevent an epidemic spreading \cite{JJ26_zhou2017numerical} \cite{wang2014asymmetrically}, for opinion formation \cite{JJE_eames2009networks} and the impact of heterogeneity and awareness \cite{scata2016impact}. Information spreading within network of parents with diseases was modeled as competing process with opinion spreading \cite{velasquez2017interacting}.

Competing processes were also explored in the context of optimal resource allocation on multilayer networks when each node can posses single process at a time \cite{wano16}. It was shown that resource diffusion in information layer can affect epidemic spreading within physical contact layer, and it changes phase transition \cite{JJ30_chen2018optimal}. Study showed that the existence of optimal resource diffusion is leading to maximized disease suppression. Also, when looking at studies in the space of opinion and decision making, we can find some interesting approaches, e.g. (i) model that investigates to what extent opinion formation and making decision processes can influence each other \cite{alla16} or (ii) model that enables to assess the consequences that propagation of the negative information may have on the adoption of the green behaviour \cite{liti18}.

\textbf{(iii) Mixed approaches.} In \textbf{46\%} of papers, mixed approaches are analysed with ability to model both competitive and collaborative processes at the same time. For example, authors in \cite{JJ2_wang2016structural} analysed coexistence of collaborating and competitive mechanism. Increase of collaboration rate increases the ability to spread the content in all layers while without collaboration layers are independent and each viral agent spread only within one layer. At the same time, due to competitive mechanism, only a single viral agent can be assigned to the single node. 

Different possibilities were analysed in \cite{JJ4_wei2016unified} including both cooperative and competing diffusion processes as well as a hybrid combination of those two. Proposed model was parameterised with increased epidemic threshold for competing spreading processes and decreased threshold for cooperative interactions. Mixed approach was based on cooperation within the first layer and competition on the second layer. It represents situation when processes within the first layer decrease spreading in the second layer, while the second layer processes reinforce the first layer activity.

Very interesting effect can be observed in the case of disease and awareness spreading where in some cases, information/awareness spreading is suppressing disease spreading, but at the same time disease spread promotes (infected node becomes aware and is able to spread information about the contagion) information/awareness spread \cite{yang2016impact} \cite{wang2016suppressing} \cite{kazh17} \cite{JJ26_zhou2017numerical}. It can be modelled as mixed model discussed in several works \cite{funk2009spread} \cite{JJ32_zheng2018interplay} \cite{JJ33_sagar2018effect} \cite{JJ3_gao2016competing} \cite{JJ8_wang2017epidemic} \cite{JJ15_gao2018dynamical} \cite{JJ20_huang2018global}. Interaction between awareness and epidemics spreading can be observed and modelled with the use of activity driven model \cite{gule16}, risk perceptions \cite{JJ29_zhou2018immunizations}, or approach that also takes into account individual behaviour where a node can decide whether to communicate with nodes that are sources of information or disease \cite{faji16}.

Similar scenario to 'disease supporting spread of awareness and awareness suppressing disease' was also explored in the case of disease and immunization where spread of diseases is modelled over multilayer structure and immunization strategies can enhance them or impair \cite{JJ29_zhou2018immunizations} \cite{JJ28_massaro2014epidemic}. Immunization can compete with epidemics but, at the same time, epidemics can enhance the dynamics of immunization \cite{JJA_jo2006immunization}. Similar mixed interaction can be observed for awareness when vaccination is used in \cite{lich17} \cite{liwa16}. 

\textbf{(iv) No interaction.} In \textbf{7\%} of papers there is no interactions between processes. In \cite{czaplicka2016competition} the processes are competing, but they spread the same content, while in \cite{joto17} contagions neither compete nor collaborate and every node can be infected by an arbitrary number of contagions.

\subsubsection{Spread timeline}
\begin{table*}
\begin{center}
\begin{tabular}{l|rp{12cm}} \hline
Timeline & No. of papers& References \\ \hline
Sequential & 1 (2\%) & \cite{funk2010interacting}\\
Concurrent & 55 (94\%) & 
\cite{funk2009spread}\cite{JJA_jo2006immunization}\cite{weva13} \cite{wei2012competing}\cite{scata2018quantifying}\cite{guo2016role}  \cite{JJ8_wang2017epidemic} \cite{sahneh2014competitive}\cite{joto17} \cite{dabo17} \cite{wano16} \cite{zhxu18} \cite{wuxi18}\cite{marceau2011modeling} \cite{sanz2014dynamics} \cite{sahneh2013may} \cite{azimi2016cooperative} \cite{JJ4_wei2016unified}\cite{JJ13_wei2018cooperative} \cite{JJ29_zhou2018immunizations}\cite{grgo14}  \cite{jusa18} \cite{faji16} \cite{gule16} \cite{lich17} \cite{liwa16}\cite{kazh17} \cite{grgo13}\cite{czaplicka2016competition} \cite{wang2014asymmetrically} \cite{velasquez2017interacting} \cite{wang2016suppressing} \cite{scata2016impact} \cite{nicosia2017collective} \cite{nie2017impact} \cite{pan2018impact} \cite{pan2018impact2} \cite{xia2019new} \cite{zhang2017epidemic} \cite{liu2016community} \cite{zhang2018new} \cite{yang2016impact}\cite{JJ3_gao2016competing}\cite{JJ15_gao2018dynamical} \cite{JJ20_huang2018global} \cite{JJ21_zang2018effects} \cite{JJ26_zhou2017numerical}\cite{JJ28_massaro2014epidemic} \cite{JJ30_chen2018optimal}\cite{JJ32_zheng2018interplay}\cite{JJ35_guo2015two}\cite{JJE_eames2009networks}\cite{liti18}\cite{alla16} \cite{JJ34_chang2018co}  \\
Overlapping & 1 (2\%) & \cite{marceau2011modeling} \\
Not specified & 1 (2\%) & \cite{JJ2_wang2016structural} \\ \hline
\end{tabular}
\end{center}
\caption{The timeline of both processes.}
\label{tab:spreadingtimeline}
\end{table*}
The vast majority of research proposed models where both spreads are concurrent. There is very limited research into more realistic modelling where the awareness/immunisation reaction is delayed as in \cite{marceau2011modeling}. There is also small sub-area where first the protection against the virus is spread, and after that, the virus spreads over network that undergone some immunisation process \cite{funk2010interacting}. Other than those couple of studies, once again we can see a tendency to simplify the process and avoid the complexity connected with delaying some of the processes as it is in the real-world.

\subsection{Where? and How?}

\begin{table*}
\begin{center}
\begin{tabular}{p{1.9cm}|p{2cm}|p{1.9cm}|p{1.8cm}|p{1.8cm}|p{1.6cm}|p{1.6cm}|p{1.6cm}}
\hline
Network models &SIS SIS&Other&SIR SIR&SI1SI2S SI1SI2S&SIS SIR&SIS TM&SIR SIRV \\ \hline
Scale Free \newline Scale Free 
&
\cite{sanz2014dynamics}\cite{JJ4_wei2016unified}\cite{grgo14}\cite{jusa18}\cite{faji16}\cite{kazh17} \cite{grgo13}\cite{nie2017impact}\cite{pan2018impact}\cite{zhang2017epidemic}
\cite{zhang2018new}\cite{JJ20_huang2018global}\cite{JJ26_zhou2017numerical}\cite{JJ28_massaro2014epidemic}\cite{JJ30_chen2018optimal}\cite{JJ33_sagar2018effect}\cite{liti18}
& 
\cite{liu2016community}\cite{JJ21_zang2018effects}\cite{JJ35_guo2015two}
& 
\cite{zhxu18}\cite{sanz2014dynamics}\cite{JJ29_zhou2018immunizations}\cite{scata2016impact}
& 
& 
\cite{xia2019new}\cite{JJ32_zheng2018interplay}
& 
\cite{pan2018impact2}
& 
\cite{lich17} 
\\ \hline
Poisson (ER) \newline Poisson (ER) 
&
\cite{joto17}\cite{dabo17}\cite{wuxi18}\cite{sanz2014dynamics}\cite{jusa18}\cite{JJ26_zhou2017numerical}\cite{JJ28_massaro2014epidemic}
& 
\cite{JJ2_wang2016structural}\cite{wang2016suppressing}\cite{JJ21_zang2018effects}\cite{JJ35_guo2015two}\cite{JJE_eames2009networks}
& 

\cite{zhxu18}\cite{marceau2011modeling}\cite{sanz2014dynamics}\cite{azimi2016cooperative}
\cite{JJ29_zhou2018immunizations}\cite{funk2010interacting}
& 
\cite{wano16}
& 
\cite{yang2016impact}
& 
\cite{guo2016role}\cite{czaplicka2016competition}
& 
\\ \hline
Poisson (ER) \newline Scale Free
&
\cite{JJ8_wang2017epidemic}\cite{jusa18}\cite{JJ15_gao2018dynamical}
& 
\cite{nicosia2017collective}
&
\cite{marceau2011modeling}
& 
\cite{weva13}\cite{wei2012competing}\cite{sahneh2014competitive}\cite{sahneh2013may}
& 
\cite{xia2019new}
& 
\cite{guo2016role}
& 
\cite{liwa16}\cite{wang2014asymmetrically}
\\ \hline
Regular \newline Regular
&
\cite{jusa18}\cite{JJ26_zhou2017numerical}\cite{JJ28_massaro2014epidemic}
& 
\cite{funk2009spread}\cite{alla16}\cite{JJ34_chang2018co}
& 
& 
& 
& 
& 
\\ \hline
Small World \newline Small World
&
\cite{JJ13_wei2018cooperative}\cite{pan2018impact}\cite{liti18}
& 
& 
\cite{zhxu18}\cite{JJ29_zhou2018immunizations}
& 
& 
& 
& 
\\ \hline
Other \newline Other
&
& 
\cite{velasquez2017interacting}\cite{JJ3_gao2016competing}
& 
& 
\cite{weva13}\cite{wei2012competing}
& 
& 
& 
\\ \hline
Small World \newline Scale Free
&
&
& 
\cite{JJA_jo2006immunization}
& 
& 
& 
& 
\\ \hline
Poisson (ER) \newline Exponential
&
\cite{jusa18}
& 
& 
& 
& 
& 
& 
\\ \hline
Poisson (ER) \newline Regular
&
\cite{jusa18}
& 
& 
& 
& 
& 
& 
\\ \hline
Exponential \newline Exponential
&
\cite{jusa18}
& 
& 
& 
& 
& 
& 
\\ \hline
Exponential \newline Scale Free
&
\cite{jusa18}
& 
& 
& 
& 
& 
& 
\\ \hline
Exponential \newline Regular
&
\cite{jusa18}
& 
& 
& 
& 
& 
& 
\\ \hline
Scale Free \newline Regular
&
\cite{jusa18}
& 
& 
& 
& 
& 
& 
\\ \hline
Scale Free \newline Other
&
\cite{gule16} 
& 
& 
& 
& 
& 
& 
\\ \hline
\end{tabular}
\end{center}
\caption{Spreading models used for network types combinations}
\label{tab:NetworksAndSpreadingModelsRef}
\end{table*}

The field of multispread in multilayer networks is big. There is no question about it. There is a number of variables and their ranges that can be considered, and many of them are still uncharted territory. The characteristics common to all reviewed papers are (i) the spread model and (ii) network topology as they are cornerstones of any spread analysis over the network. 

\begin{figure*}[ht]
\includegraphics[width=\textwidth]{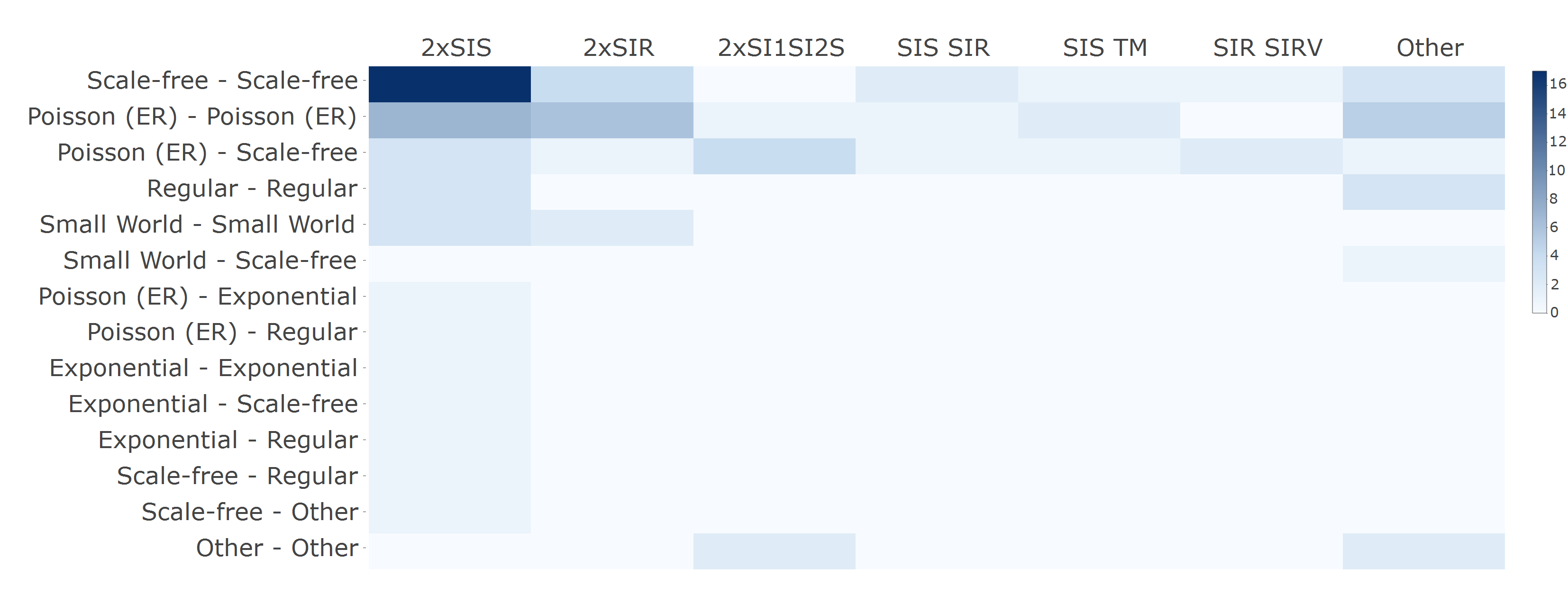}
\caption{Heatmap representation of papers published number for specific spreading models and network types.}
\label{fig:heatmap1}
\end{figure*}

In Table~\ref{tab:NetworksAndSpreadingModelsRef} we present which network and spread models were most commonly used together. We look into two layer and two-spread scenario as it is the most often explored setting (see Table~\ref{tab:NumberOfLayers} where it is shown that 95\% of studies used two layer networks in their experiments). As with other analyses presented above, we can see the tendency to simplify the complex problem of multispread over multilayer network. In most cases, researchers look at networks that follow the same model and the same type of process spreads on both layers. The biggest number of studies look into multilayer network with both layers being Scale Free networks and SIS process spreading over each layer. The popular settings are also: (i) Poisson-Poisson network with SIS-SIS spread; (ii) Poisson-Poisson network with SIR-SIR spread; (iii) Small World-Small World network with SIS-SIS spread, and (iv) Poisson-Scale Free network with SI1DI2S-SI1SI2S spread. Please note that only the last setting has network composed of two layers that are generated using two different models. The others use traditional network models with the basic spread models. This shows that we are yet to explore and understand the multispread phenomenon over more complex, more realistic structures. Fig. ~\ref{fig:heatmap1} shows heatmap with representation of number of papers published for specific spreading models and network types combinations.

\subsection{Why it spreads in this way? - Knowledge Synthesis} 
\label{sec:why}

There are several characteristics that have been empirically investigated and tested. Those analyses resulted in a variety of conclusions that show a wide variety of experiments conducted. When it comes to analysis and comparison of findings from different studies, it is not possible as every study uses its own settings. Thus the results are not directly comparable. In this situation, we decided to present the main conclusions and organise them in a way that is consistent with what we presented in Sections~\ref{sec:where}, and~\ref{sec:how}.
Thus, we start with studies that looked into how the structure and characteristics of multilayer network influence the spread. We follow with discussion about how features of multispread influence the propagation process. Finally, we present some other, significant conclusions, not directly related to the previous points.

\subsubsection{Influence of multilayer network's features on the spreading process}

\begin{enumerate}

\item{\textbf{Networks diversity}}
Diversity of a network and its layers can be expressed in many ways, but the most commonly analysed that we discuss below are (i) node degree correlation and (ii) overlap between network layers. What research shows is that epidemics on low diversity networks depends on one layer while performance in high diversity networks is more dependent on collaboration \cite{JJ2_wang2016structural}. We explore that statement in the context of the reviewed literature.
\\
\textbf{* Node degree correlation}
\\
\noindent\fbox{%
 \parbox{0.41\textwidth}{%
\textbf{FINDINGS:}\\
* \textbf{Positive Correlation}: One (stronger) spread is more effective than another (competing, multiplex scenario). Additionally when heterogeneity (variance in the degree distribution) is higher one of the spreads is even more effective. \\
* \textbf{Negative Correlation}: Coexistence region for processes is larger (competing, multiplex scenario)\\
* \textbf{Low impact of correlation when spread cooperate}\\
* \textbf{Ergo, low diversity means that one spread influences another one more than when diversity is high}
 }%
}
\\\\
Results from reviewed studies show that in the case of competing scenario, where disease spreads on one layer and intervention on another, the positive degree correlation between networks' layers increases the efficiency of the intervention \cite{marceau2011modeling} (disease spreads over Poisson or SF network; intervention spreads over SF network, SIR model for both layers), \cite{funk2010interacting} (Random-Random; SIR-SIR). Similarly, in \cite{nie2017impact} authors show that if nodes degree in their two layer (SF-SF) multiplex network are correlated the awareness spread (UAU) has higher suppressing effect to the epidemic spreading (SIS). Additionally, protective effect is stronger if there is more significant variance in the degree distribution (higher heterogeneity) \cite{funk2010interacting}. For no correlation of nodes degree or negative one, increasing the heterogeneity makes it more and more challenging to contain the second spread which is epidemic that follows awareness campaign (second spread is faster) \cite{funk2010interacting}.

Positive degree correlation causes that it is easy to remove the virus from the system in a scenario where two viruses compete (survival threshold is larger for positively correlated layers) \cite{sahneh2014competitive} (Random-SF; SI1I2S-SI1I2S).
On the other hand, the negative correlation, for two competing viruses, makes survival for one virus easier but, at the same time, it proves to be more challenging to remove the other contagion entirely \cite{sahneh2014competitive}. For negatively correlated layers the coexistence region, where both viruses exist, is bigger \cite{sahneh2014competitive}.

Similar findings were reported for both competing scenarios: (i) virus-virus and (ii) virus-protection/awareness. This shows that from the perspective of design both are alike. Indeed, SIS spread model for virus is the same as UAU model for awareness spread. 

When analysed spreads cooperate/support each other, researchers in \cite{JJ13_wei2018cooperative} concluded that the epidemic threshold is not significantly influenced by the interlayer degree correlation and at the same time they noticed that large interlayer degree correlation is resulting in lower prevalence. Experiments based on SIS spreading model were performed for randomly-correlated homogeneous network with two layers based on Small World,  for correlated heterogeneous networks with Scale Free model on each layer and the same model for uncorrelated networks.   
\\
\\
\textbf{* Network overlap}
\\
\noindent\fbox{%
 \parbox{0.41\textwidth}{%
\textbf{FINDINGS:}\\
* \textbf{Higher overlap} between layers boosts one of the spreads (dominant one) -- e.g. awareness can  spread faster\\
* \textbf{Small overlap} means that awareness cannot influence the disease layer -- hard to stop the disease\\
* \textbf{Ergo, low diversity means that one spread influences another one more than when diversity is high}\\
* Role of overlap on awareness spreading is moderated for low and high propagation probability
 }%
}
\\\\
The influence of layers overlap on the diffusion processes is one of the popular elements investigated in the reviewed field. It has been found that, in the competing and mixed scenarios (e.g. awareness vs disease spread), higher overlap between network layers facilitates the invasion of the undesirable process (SIR-SIR; SF/Poisson-SF; spread of two competing viral agents or spread of disease vs spread of intervention) \cite{marceau2011modeling}. It can also amplify the effect of awareness spread  \cite{funk2009spread} (regular graphs on both layers, SIR-awareness enhanced SIR), \cite{guo2016role} (SF--SF two layer multiplex, TM-SIS), \cite{wang2014asymmetrically} (SF-ER two layer multiplex, SIRV-SIR), and \cite{yang2016impact} (ER-ER two layer multiplex, UAU-SIR). In general, both strands of research show the same: high overlap contributes to the faster spread of a stronger (dominant) process. The higher overlap between the layer where disease spreads and layer in which people exchange information and communicate also helps to enhance the effect of locally spreading awareness (defined as the behavioural response arising in the region where disease outbreak). It is especially visible in networks which have high clustering (awareness vs epidemic) \cite{funk2009spread}. In such a case, the disease, as long as its infection rate is below threshold, can be completely stopped. The importance of the local risk of infection information received from neighbours in information layer that reduces the node susceptibility in contact layer was also investigated and emphasized in \cite{wuxi18}. Authors introduce the individual awareness element (dependent on the number of infected nodes) that is able to change the infection rate. The results show that in the setup with two layer ER network where both processes follow SIS model with individual awareness, only the information from node's neighbourhood that overlap in two layers can have effect on the epidemic threshold. The higher the overlap, the higher the epidemic threshold.

Another study with the use of uncorrelated two layer network generated with configurational model and SIS based spreading looks at resource diffusion strategies (for example information campaign budget). It shows that they can be adjusted to different levels of interlayer correlations between information and pathogen layer with ability to maximally suppress diseases above thresholds with maximum values \cite{JJ30_chen2018optimal}.

In \cite{sahneh2013may} authors are using SI1SI2S model to investigate the interaction between two competing viruses on two layer (ER-SF) multiplex network and show that it is easier to remove a virus from the system when the network layers are positively correlated. On the other hand, when they are negatively correlated it is much harder to remove a virus. Authors showed it using both analytical and numerical methods.

It is also confirmed by \cite{JJ28_massaro2014epidemic} where authors with the use SIS model on Scale Free, Small World and Random networks show that increasing differences between information (where awareness spreads) and contact network (where disease spreads), makes the task of stopping epidemics more difficult. In real systems it can be observed for diseases developing in regions with low access to the Internet meaning that overlap between those networks is small \cite{JJ28_massaro2014epidemic}.

There is also some research that the percentage of overlapped links does not influence the spread but that the percentage of susceptible nodes can have big effect \cite{zhxu18}. Authors tested it on the case of two interacting diseases where each contagion spreads, using modified SIR model, over one layer but both layers follow the same network model for a given experimental setup (Random, Small World or Scale Free). However, as authors point out, this may be the effect of the experiment setup where interplay between two contagions is nodes based. It means that communication takes place if node in state $S$ for one disease is in the $I$ or $R$ state for the other disease.

Overlap can be computed for networks with different nodes. For example \cite{JJE_eames2009networks} analyses the network of communication and sharing opinion parents and associated network of children where edges between them can be similar like in the parents network. And the similarity is the measure of overlap. In this case parent's opinion has stronger effect on suppressing disease spread between children when network of physical contacts between children has higher overlap with information networks of parents. Authors generated parents' network by adding random links with probability dependent on separation distance. Kids network was simplified and generated with assumption that only one kid can be assigned to parents and the same connections exist. Apart from used algorithm some fraction of links was added randomly to both networks. Disease transmission and recovery rate for kids was used while for the parents opinion formation process was modeled. 

Apart from investigating the influence overlap, \cite{jusa18} developed a toolbox of algorithms to that enables to generate two layer multiplex network with given node degree distributions and with a predefined overlap coefficient (Jaccard index).
\\
\item{\textbf{Edges between layers and spread switching layers}}
\\
\noindent\fbox{%
 \parbox{0.41\textwidth}{%
\textbf{FINDINGS:}\\
* \textbf{The more interlayer connections} in the multilayer networks, \textbf{the easier} for the spreading process \textbf{to affect all nodes}. \\
* \textbf{The spread with the higher probability of switching the layers} has an advantage over the other one as it \textbf{can easier spill over the other layer}}%
}
\\\\
In \cite{czaplicka2016competition} authors analyse how the number and distribution of interlayer links affects spreading (phase transitions/epidemic threshold) of the same disease using threshold model and SIS model on two layer (ER-ER) network. They found that "in the threshold layer the critical value of the threshold increases with the interlayer connectivity, whereas in the case of an isolated single network it would decrease with average connectivity"  \cite{czaplicka2016competition}. If the threshold in Threshold Model is below the critical value all nodes become adopters and if the threshold is above the critical threshold only the initial nodes that belonged to the adopters set remain as adopters and in consequence the spread of adoption does not happen.
In SIS layer, the interlayer coupling seems to be the reason for the transition between the situations where we observe small and large percentage of nodes that become adopters.   
a new transition between regions of low and large number of adopters appears to be caused by the interlayer coupling. 

Also, authors in \cite{alla16} (degree-regular random networks on both layers; M-model for opinion dynamics and Abrams-Strogatz model for decision dynamics) consider pairwise connections between layers. In their model each vertex is connected to a single, randomly selected vertex in the other layer. However, this is treated as an element of the model and influence of those links between layers is not investigated and left for future work.

Interaction between the layers can be modelled not only by adding edges between layers but simply by enabling the spread to switch the layers. And while in most papers spread of different medium takes place only on separate layers, some works assumed that content can spread through all layers. Spread switching layers was observed for memes \cite{wei2012competing} and \cite{weva13} (SI1I2S spreads on both layers; different sythetic network structures analysed) and for viral agent using all edges in all layers with simple SI model and networks with a Poisson degree distribution \cite{JJ2_wang2016structural}. When layer switching is possible infection of node can be initiated by neighbours on different layers \cite{JJ13_wei2018cooperative} as well as recovery is possible at all layers with different probabilities \cite{JJA_jo2006immunization}. 
In \cite{wei2012competing} and \cite{weva13} authors show that the spread with the higher probability of switching the layers has an advantage over the other one as it can easier spill over the other layer. This confirms the finding from \cite{czaplicka2016competition} -- the more interaction between layers or the higher the probability of jumping between the layers the easier for a spread to affect all the nodes.
\\ 
\item{\textbf{External influence}}
\\
\noindent\fbox{%
 \parbox{0.41\textwidth}{%
\textbf{FINDINGS:}\\
* \textbf{Greater external influence} on increasing awareness -- the epidemic threshold is larger \\
* \textbf{Greater external influence} on increasing awareness -- the onset of the epidemic is delayed
 }%
}
\\\\
In section \ref{subsec:ExternalInfluence} we discussed how the external environment can be taken into account. In majority of cases it is "mass media" like approach where e.g. certain percentage of population becomes aware of a disease. It can be e.g. (i) a random process, where external node representing the mass media connected to all nodes in the information layer, regularly and randomly sends information about the disease \cite{grgo14}, (ii) model that assumes that probability of being aware depends on the global percentage of aware individuals \cite{JJ21_zang2018effects}, or (iii) approach  where mass media influences the awareness level depending on how many people is infected \cite{kazh17}.

In general, as one can expect, the tendency is that the bigger mass media effect, the onset of the epidemic is delayed \cite{grgo14} (SIS-UAU; layers being power law networks), \cite{xia2019new} (SF-SF two layer multiplex, UAU-SIS), \cite{JJ21_zang2018effects} (two layer Scale Free network and SIS–UAU model). So, the transmitting information about the disease is critical and highly influence the final outcome of the epidemics helping to control the spread.

Although some work has been done in considering the external environment, it only accounts for scenario where system awareness about the disease is increased by making some of the nodes arbitrary aware of the disease.

\end{enumerate}

\subsubsection{Influence of spread characteristics on the spreading process}
\begin{enumerate}
\item \textbf{Interaction between processes}
\\
\noindent\fbox{%
 \parbox{0.41\textwidth}{%
\textbf{FINDINGS:}\\
* \textbf{Competing processes} -- epidemic thresholds increase \\
* \textbf{Cooperative processes} -- epidemic thresholds decrease\\
* \textbf{Interacting processes} -- epidemic thresholds can be increased or decreased by activity of processes in other layers when compared to isolated layers
 }%
}
\\\\
One of the most investigated elements in the mutlispread scenario is how the type of the interaction between processes influences the properties of the spread (e.g. epidemic threshold). Research shows that in the case of multiplex networks, epidemic thresholds can be increased or decreased depending on the character of the spreading processes and nature of the interaction between them (e.g. competing, cooperative, mixed) when compared to isolated layers \cite{JJ4_wei2016unified}. In various studies analysed in this review, we can see that, in general, competing processes increase epidemic thresholds while two cooperative ones decrease epidemic thresholds. 
Below, we organise the description of the interactions between processes according to 'what' spreads over the network as this is the commonly used taxonomy in the literature and in the same the most intuitive way of presenting different types of interactions between processes.
\\\\
\textbf{Disease-Disease spreads}. Sanz et al. \cite{sanz2014dynamics} observe the interaction between two competing diseases on two layer multiplex networks (Scale Free--Scale Free and ER-ER) using SIS-SIS and SIR-SIR models. "The results show that there are regions of the parameter space in which the onset of a disease's outbreak is conditioned to the prevalence levels of the other disease" \cite{sanz2014dynamics}. Moreover, for the SIS-SIS scheme, they found out that "under certain circumstances, finite and not vanishing epidemic thresholds are found even for Scale Free networks" \cite{sanz2014dynamics}. Finally for SIS-SIS the secondary threshold is different than for SIR-SIR scheme and this is a consequence of how both diseases interact with each other. 

While Sanz has been analysing competing diseases, the Azimi-Tafreshi in \cite{azimi2016cooperative} has been evaluating how the presence of one contagion can boost the diffusion of the other infection when they support each other. They found out that "cooperation of two diseases decreases the network's robustness against propagation of both diseases, such that the epidemic threshold is shifted to smaller values" \cite{azimi2016cooperative}. The low cooperativity means that the co-infected cluster (where all nodes are infected with both diseases) "emerges continuously, however, increasing the strength of cooperation, the type of phase transition changes to hybrid and tricritical point emerges" \cite{azimi2016cooperative}. The experiments and analytical evaluation were performed for two layer (ER-ER) multiplex network with two SIR epidemic models. The fact that two diseases helping each other weaken the human immune system is also confirmed by \cite{JJ8_wang2017epidemic} with the use of different time scales for awareness and epidemic spreading. Experiments were focused on two layer ER-ER and Scale Free--Scale Free networks with the use of SIS-UAU models. However, in this case, authors emphasize that after recovery it is more difficult to spread the disease again.

In \cite{sahneh2013may} authors are using SI1SI2S model to investigate the interaction between two competing viruses on two layer (ER-SF) multiplex network and are able do identify survival threshold and winning threshold i.e. the conditions under which two compeering viruses are (i) able to coexist and (ii) will lead to extinction of one of them.
\\\\
\textbf{Epidemic-Awareness spreads}. All reviewed papers agree that information spreading slows down or even stop the disease spreading however they differ on some aspects of their models, experimental setup or elements considered. 

In \cite{wang2014asymmetrically} authors analyse interplay between disease (SIRV) and information spreading (SIR) on two layer (SF-ER) multiplex network. They show that "epidemic outbreak on the contact layer can induce an outbreak on the communication layer, and information spreading can effectively raise the epidemic threshold, and when structural correlation exists between the two layers (layer overlap), the information threshold remains unchanged but the epidemic threshold can be enhanced, making the contact layer more resilient to epidemic outbreak" \cite{wang2014asymmetrically}. Similar research was done in \cite{wang2016suppressing} which shows that an optimal information transmission rate can be identified for which the infection diffusion can be effectively suppressed. The only difference was that information was spreading with SIS model and that both layers were ER networks. Scata et al. had similar findings i.e. they show that  awareness spreading (UAF) can delay the outbreak of the infection spread (SIR) and is capable of strengthening of the node's resilience in two layer (SF-SF) multiplex network \cite{scata2016impact} and three layer (SF-SF-SF) weighted multiplex network \cite{scata2018quantifying}.The inhibition effect of information propagation (SIR model) on the spread of the disease (SIRV model) has been also shown by \cite{lich17} when both layers are Scale Free networks. But they also noted that the information spread becomes less effective when the vaccination probability (as one of the states in SIRV model) is positively correlated with individual's degree on the contact layer (where disease spreads). In such a case only high degree nodes adopt the vaccination behaviour. In a situation when the vaccination probability is negatively correlated with node's degree then the nodes with low degree will be vaccinated and, as the experiments were run on Scale Free networks, this results in a state where majority of the individuals are vaccinated and in turn the information spread will be even more effective. 

In \cite{pan2018impact} researchers study how local and global information (UAU) affects epidemic (SIS). The results show that, in two layer (SF-SF) multiplex network, the percentage of infected individuals can decrease due to contact-based precautions and this, in turn, causes that the epidemic threshold increases. 
On the other hand, "prevalence based precautions, regardless of local or global information, can only decrease the epidemic prevalence" \cite{pan2018impact}. Additionally, they found that the altruistic behaviours of infected people are a good way in suppressing the spread of the disease what is in line with previous studies. 
Guo et al. \cite{guo2016role} challenge previous assumptions that the awareness spreading affect all nodes in the same way and explore the how the  heterogeneity of nodes influences diffusion of disease. Using the k-core and the degree measures, they clustered the individuals into groups. Each node in a given groups is characterised by the same local awareness threshold or probability of being infected. Next they observe interplay between between epidemic (SIS) and awareness (TM) on two layer (SF-SF, ER-ER, SF-ER) multiplex network and find that change in the nodes heterogeneity significantly affects epidemic threshold and final number increasing or decreasing it.

Pan et al. \cite{pan2018impact2} takes it a step further and proposes three types of heterogeneity to their model: (i) the heterogeneity resulting from the variety of types of responses to epidemic outbreaks that people can come up with and  the influence heterogeneity that is present in (ii) the disease layer and/or (iii) in the information layer. The main goal of their research is to understand the impact of the different heterogeneity types on the interaction between  disease and awareness. They perform the simulation in similar seating as in \cite{guo2016role}, i.e., epidermic - SIS, awareness - TM, two layer (SF-SF) multiplex network, and confirm that changing heterogeneity level affects epidemic threshold. The individual behaviour, where each person can make independent decision whether to contact with those who are the sources of information or disease, is also part of the spread of disease-awareness (SIS-UAU) over multiplex network (SF-SF) studies in \cite{faji16}. Similarly to other research, authors found that more reasonable people reactions ad behaviour together with the propagation of the information is an effective way to reduce the epidemic spread.

In \cite{liu2016community} authors study how community structure affect the coupled disease (SIR) awareness (IC) spreading in two (SF-SF) layer multiplex network with and without community structure, and find out that the diffusion of the epidemic is significantly impacted by two elements: (i) the number of groups and (ii) the level of the overlap between the groups on different network layers. Another interesting work in this area was presented in \cite{zhang2018new}. The results are similar as above i.e. awareness spreading can slow down or even stop disease spreading, but their experimental setup is unique. They investigate two layer (SF-SF) multilayer network where one layer represent communication between parents where the awareness (UAU) about the disease among children is spreading. The second layer represent physical contacts between children where disease (SIS) is spreading. 

Another study identifies capacities and awareness diffusion and self protection \cite{JJ3_gao2016competing}. Presented approach takes into account interplay between epidemic spreading processes and self-protection ability of aware nodes represented by lower infectivity. Study is based on SIR and SIRI models and two network layers with the same nodes but different topologies. Competing processes where modeled with microscopic Markov chains and self protection influencing infectivity. Results show that increased capacities increase epidemic thresholds and as a result the outbreak size is smaller. Authors investigated targeted and random immunization what showed much better performance of targeted immunization focused on high degree nodes than random approach. However, epidemic threshold and outbreak size were not dependent on self-awareness assigned to nodes within information layers after infection in physical contact layer. Also, \cite{kazh17} confirms that self-awareness cannot alter the epidemic threshold on two layer Scale Free network (SF-SF) where SIS spread model is deployed on both layers.

The research also shows that the awareness spreading (following SIS or threshold model) can both increase the threshold of the epidemic (SIS spread over static SF network) and reduce the percentage of infected nodes \cite{gule16}. What is more important, authors also looked into how the structure of the information network (generated using activity driven model) changes over time  and they concluded that those changes causes that the awareness spread diminishes and this has a direct effect on the epidemic threshold. 

Not only the epidemic threshold was analysed but some research -- \cite{grgo13} \cite{grgo14} -- also discovers the existence of the metacritical point where the awareness spread (UAU) can effectively control the epidemics outbreak (SIS) on two layer (Scale Free--Scale Free) multiplex network.

Apart from typical epidemic thresholds modeling, the local awareness ratio defined as a proportion of aware neighbours to not aware ones can be utilised in the process modeling interplay between diffusion of awareness and epidemics \cite{JJ35_guo2015two}. Experiments based on two layers SF-SF networks, microscopic Markov chains and awareness controlled spreading (LACS) showed that if awareness ratio is increased the outbreak of epidemics is accelerated.
\\\\
\textbf{Epidemic-Opinion spreads}. In \cite{velasquez2017interacting} authors are investigating how diffusion of epidemics and formation of opinion impacts each other in two layer multiplex network (Poison-Poison). For the disease spreading and Contact Process (asynchronous SIS model) was used and for influence spreading the Voter model. They found that the opinion dynamics influence statistical properties of spread of the disease. "The most important is that the smooth (continuous) transition from a healthy to an endemic phase observed in the contact process, as the infection probability increases beyond a threshold, becomes discontinuous in the two layer network. Also, an endemic-healthy discontinuous transition is found when the coupling overcomes a threshold value. Furthermore, they found out that the disease dynamics delays the opinion consensus" \cite{velasquez2017interacting}. Similar situation was modelled in \cite{JJE_eames2009networks} for parents and their kids network. Epidemic transmission within physical contact network created for kids was parallel to opinion formation within parents network. Influence of information network on disease spreading process was increasing together with the increase of similarity between networks of parents and children. The higher overlap between parents clusters not supporting vaccination was resulting in emergence of children clusters without protection. Support for vaccination is increasing together with th number of infections in parents' neighbourhood.      
\\

 \item \textbf{Spread suppression and prevention techniques}
\\
\\
\noindent\fbox{%
 \parbox{0.41\textwidth}{%
\textbf{FINDINGS:}\\
* \textbf{Suppression} -- removing the nodes with the highest node degree is the most effective topology-based suppression technique\\
* \textbf{Suppression} -- appropriate time scale for suppression mechanism is a key in finding the optimal way to contain the disease\\
* \textbf{Prevention} -- the stronger the immunisation effect, the smaller number of infected nodes
 }%
}
\\
Understanding how multiple processes, through competition/cooperation or interaction mechanisms (presented above), influence the state of the system is a key for developing spread suppression and prevention techniques. Those techniques, in turn, enable to control the spread. Both, prevention and suppression have the same goal but the former one is pro-active and the latter reactive behaviour. Thus, we consider them together.

Some researchers made attempts to assess the impact of different suppression mechanisms by (i) suppressing both spreads at the same time or (ii) unilaterally suppressing a one of them and, at the same time, not affecting the other spread \cite{weva13} \cite{wei2012competing} (SI1I2S spreads on both layers; different synthetic and real-world networks analysed). All the analyses, in their case, are based on the eigenvalues of system matrices $S = (1 - \gamma)\cdot I + A$ for adjacency matrices $A_1$ and $A_2$ representing layers one and two of a network, where $\gamma$ is a meme persistence and $\beta$ is a meme strength. Authors understand the suppression as pushing $\lambda$ for one (unilateral) or both (concurrent) memes below one. They used variety of suppression techniques, including (i) Random, (ii) Acquaintance (acquaintance immunization, randomly select a node and using random approach remove one of nodes connected to it), (iii) Greedy (delete a vertex that results in the largest drop in the eigenvalue of the system), (iv) Max Degree (remove node with the highest degree), or (v) Social Hierarchy. Authors showed that, in order to control the spread, Max Degree approach is the most promising approach. 

Apart from designing targeted suppression techniques, where nodes can be removed from the network, any scenario where the awareness spreads over the network, in order to limit the outbreak of another spread, can be seen as a suppression technique. It is reactive action, we inform people about the disease because there were some reported cases of it. However, this topic is covered in detail above, please see (1) Interaction between processes -- "Epidemic-Awareness spreads" section, thus it is not further analysed in here. On the other end of the spectrum, but similar in functioning, is the negative information spread to suppress some positive behaviour. Author in \cite{liti18} shows that negative messages about the green behaviour will reduce the adoption level of green behaviour. This effect is visible for both analysed set-ups: (i) Small World networks for both layers and (ii) Scale Free networks for both layers. In both cases SIS-like spread models are considered for both layers.

One of the elements that can play the role in how fast the outbreak can be contained are the time factors related to speed of (i) information/awareness spreading and forgetting as well as (ii) epidemics spreading and recovering with the SIS-UAU model within two layer ER-ER, SF-SF and real networks \cite{JJ8_wang2017epidemic}. Research showed that the epidemic mitigation effect is related more to the time scale of the spread of information rather than to the time scale of the diffusion of the disease. Authors in \cite{JJ8_wang2017epidemic} prove that selection of optimal mitigation strategy is possible when relative time scale for information propagation is derived from awareness spreading rate. Proper time scale selection is resulting in low fraction of infected nodes. Study also shows that too fast spreading of information reduces mitigation effect and optimal time scale for information spreading is not necessary infinite because performance is also dependent on time from infection to aware state. Proposed approach covers real situations when people receive information within different contact networks, use phone communication with different frequencies, are forgetting information with different speed depending on source or form of message. 

Researchers also looked into the prevention area where the action is taken before the spread outbreak. And although there is a fine line between suppression and prevention, we classify here immunisation as a prevention technique. The main reason for that is that in the real-world the immunisation is treated as preventive action -- we do not have to have epidemic outbreak to provide immunisation; we do it to prevent the outbreak to occur. 

Research shows that the degree of immunization (regulated as a certain probability), in the setting where being informed and aware of a disease does not directly translate into being entirely immune, affects the critical properties of the system. In general, the epidemic incidence decreases and epidemic threshold increases with the increase in the level of immunization~\cite{grgo14} (SF-SF, SIS-UAU), \cite{liwa16} (SF network and SIR model for communication layer, ER network and SIRV model for contact layer), or \cite{lich17} (SF network and SIR model for communication layer, SF network and SIRV model for contact layer). 

The vaccination is often modelled as one of the states in the disease spread model, e.g. SIRV (susceptible-infected-recovered-vaccinated) model \cite{liwa16} or \cite{lich17} where a given node becomes vaccinated with a certain probability that depends on e.g. (i) receiving information and the node degree \cite{liwa16} or (ii) the number of times a node received the information about the disease coupled with the social reinforcement effect \cite{lich17}. Immunization by prevention can be also performed by assigning to nodes finite or infinite time when nodes stay immune what was showed for SIS model simulated on both layers of two layer SF-SF network \cite{JJA_jo2006immunization}. Approach based on finite prevention period assumed that nodes are returning to susceptible state when prevention period finishes. For large values of infection probability prevention modeled in such way increases spread of epidemics due to ability to immediately infect nodes after they finish prevention period. In another study, experiment based on random networks with one layer representing parents and second layer representing children, processes simulated with opinion formation for information layer and transmission rate for disease spreading, showed that non-overlapping links can help to decrease the dynamics of epidemics by vaccinations \cite{JJE_eames2009networks}. Different overlap between two layers was represented by the fraction of links between parents shared by children. 

\end{enumerate}

\subsubsection{Proposed frameworks}

Some of the studies proposed the overall frameworks for analysis of multispread over multilayer networks where the interaction between processes could be cooperative/supporting or competitive depending on the framework set-up \cite{joto17}, \cite{dabo17}. In those frameworks they investigate the transmission rate depending on the parameters of the spread. 
Sanz et al. \cite{sanz2014dynamics} proposed a framework where they can observe the interaction between two diseases on two layer multiplex networks using any combination of SIS and SIR models. Nicosia et al. \cite{nicosia2017collective} created more general framework to intertwine spreading processes that propagate over multilayer networks with various network structures.

Apart from dedicated frameworks new approaches are proposed as a methodological background for proposed models. For example effective degree theory based on analysis of surrounding nodes and counted number of nodes in $S$ or $I$ state was presented in \cite{JJ26_zhou2017numerical}. It delivered higher accuracy than mean field theory within epidemic propagation network for regular, random and Scale Free networks when compared with theoretical analysis.

There have been also attempts to extend the competitive multilayer processes to a set of heterogeneously parametrized processes that spread over the generalized graph layers \cite{wano16} as that enables to conduct more comprehensive approach that looks into set of different scenarios at the same time. Authors provided a first step in analyzing competitive spreading processes in multilayer environment by "finding necessary and sufficient conditions for the exponential stability for any equilibrium of the system in which one process extincts exponentially quickly and the other survives in an endemic state" \cite{wano16}. They have proposed a whole optimization regime for obtaining optimal-cost parameter distributions so they lead to the desired equilibrium, and alternative one which performs a heuristic design if the resources are limited.

Another framework is focused on modeling the spread of epidemic and information within coupled multiplex networks \cite{JJ32_zheng2018interplay}. Nodes within information network are assigned to aware and unaware classes with contact process spreading mechanics used. Microscopic Markov chain model was used to generate tree based on probabilities of switching between modeled states. Attenuation factor is used to take into account different spreading abilities of aware nodes. Model achieves results close to Monte Carlo simulations for higher attenuation while accuracy drops for attenuation equal to zero which represents situation when aware nodes are completely immune. 

Other framework designed for modelling co-evolution of epidemics and awareness spreading uses as a key parameters time variation of transmission rates \cite{JJ33_sagar2018effect}. Differential rate of transmission within epidemic layer is dependent on spreading probabilities of related nodes within information layer and the same for information layer in relation to epidemic layer. Spreading awareness and disease was modeled with the use of Monte Carlo Markov chain method. Second order linear theory was used to describe processes in continuous time in terms of coupled damped and driven oscillator equations. During simulations equilibrium state was identified when prevalence of disease and awareness takes place and transmission rates are at least equal to critical values.

\section{Future challenges and Road Map}
\label{sec:RoadMapGuidelines}
Although over the last decade, we have significantly improved our understanding of the multiple spreading processes over multilayer networks, we have still a long journey ahead of us, before we can develop models that are capable of mimicking the real-world environments. Though the challenge is big, we have to keep in mind that the more realistic representation of our world we can create, the  more significant impact our research will have. Only then our results can be translated into knowledge useful in the real-world scenarios. However, \textit{where to start?} is a tricky and at the same time a crucial  question to answer. Thus, based on the review of the current progress in the multispread over multilayer networks field, we define a set of NEEDS that should drive our future work in this area: (i) need for overarching and rigid methodology; (ii) need for diversity that builds complexity; (iii) need for data-driven approaches; (iv) need for dynamic and predictive modelling. These  NEEDS arise from the current research gaps and form a road map that suggests steps that will help to bring us from pure model-driven research to data-driven approaches mixed sensibly with model-based ones that enable to reflect real-world situations with higher precision. 

\subsection{Need for overarching and rigid methodology}
When reviewing existing literature, we were struck by a very fragmented approach to experiment setup, lack of justification for the conducted experiments and the parameters' values, and lack of comprehensive comparison of investigated models with other similar approaches. 

One of the underinvestigated elements is the approach to selecting values for the plethora of parameters, both in the network and spread models. Most of the experimental settings are set arbitrary and with no discussion that would show why these settings should be used (e.g. the type of model used to generate the network). It shows a methodological issue to be addressed -- in research, we always need to justify our decisions. Another methodological challenge we spotted is silent assumption in regards to values of some parameters -- sometimes, they are simply not reported and this means that the conducted experiments cannot be reproduced. Two potential ways to address those issues can be employed: either (i) we can use more realistic parameters that are data-driven, e.g. based on the medical literature in case of the disease spreading modelling, or (ii) to use the whole range of parameters' values and their combinations to provide comparative and comprehensive approach to conducting the experiments.

Another issue is resulting from the insufficient consideration of technical parameters within experimental setup. For example, different studies perform different number of repetitions for Monte-Carlo simulations, use different propagation probability ranges and different number of seeds within the network to initiate spreading processes. Presented papers use various analytical methods based on the bifurcation theory \cite{nopr16}, effective degree theory \cite{wei2012competing}, mean field theory \cite{alla16}\cite{dietz1967epidemics}\cite{kazh17}\cite{sash15}\cite{scata2016impact}\cite{solomonoff1951connectivity}\cite{JJ4_wei2016unified}\cite{wei2012competing}, individual-based mean-field approximation \cite{sanz2014dynamics}, percolation theory \cite{azimi2016cooperative}, Markov Chains \cite{guille2013information}\cite{gule16}, microscopic Markov Chains \cite{faji16}\cite{fuba15}\cite{funk2009spread}\cite{JJ3_gao2016competing}\cite{JJ15_gao2018dynamical}\cite{goldenberg2010survey}\cite{jusa18}\cite{newman2011structure}\cite{newman2002spread}\cite{sanz2014dynamics}\cite{JJ2_wang2016structural}, dynamic microscopic Markov Chains \cite{JJ33_sagar2018effect}\cite{sahneh2013may} and simulations with the use of agent based modeling and Monte Carlo method  \cite{alla16}\cite{fuba15} \cite{JJ3_gao2016competing} \cite{goldenberg2010survey} \cite{gule16}\cite{jusa18} \cite{kivela2014multilayer} \cite{pan2018impact} \cite{sanz2014dynamics} \cite{scata2016impact} \cite{van2009virus} \cite{JJ13_wei2018cooperative} \cite{JJ2_wang2016structural}. All the above makes the comparison of the results not feasible.

From the evaluation perspective, each paper we reviewed, performed analysis on different networks, either real or generated using standard network models but with different parameters for each research. This also contributes to the issue connected with inability to properly compare different approaches and also limited robustness of the results as there is no guarantee that given findings will hold for other structures than those tested. For example, in some cases, authors randomly added some links to one layer to create the second one but provided no justification why it was done. In such a case, there is no way to assess the robustness of the results. 

All those issues put together shows that, although there seems to be some standard in respect to what steps are needed to run experiments in the space of multispread in multilayer networks, there is no uniform and rigid methodology that would enable comprehensive and comparative analysis between different experimental settings. Thus, there is a need for an overarching scheme that would make such analysis feasible.

We  suggest,  that  for  the  sake  of  transparency  and  completeness,  all  research  in  the  field  should  (1) include the experiments on real data, (2) include the networks with various number of layers, (et least networks with  2, 3, 5 and 7 layers) and networks with increasing number of nodes/edges (100, 10,000, 100,000 and 1,000,000 of nodes/edges), (3) enable repeating the experiments. In particular each paper should be published together with (i) proper experiment setup including justification for the arbitrary values selected for each parameter,  (ii)  developed  code  for  simulations  and experiments, (iii) networks used in the experiments (both artificially generated and build based on real data), and (iv) all results of analysis including the step-by-step simulation outcomes. It this way we will contribute to open science and make it more accessible. 

\subsection{Need for diversity that builds complexity}
While all reviewed works discuss multispread in multilayer networks, in the majority of studies, researchers focus on only two layer (mostly multiplex) networks and two processes. While this fits the problem description, it is the simplest of possible scenarios. As the real-world interactive spreading processes over multilayer networks are much more complex, there is a great need for systematic work towards understanding this complexity, and this involves acknowledging and accounting for the huge diversity in this space. Only then, we will be able to push further the boundaries of our knowledge and develop approaches that will generate impact in the real-world. 

Investigated structures and spreading processes should be diversified and take into account a bigger range of scenarios. Currently, though the research space is big, most of the research focuses on random, Scale Free, Small World networks with the use of synthetic datasets based on theoretical models. It creates space for usage of other models and the need for real data collection is growing to better address real-world challenges. Only few papers were using real datasets. Additionally, experiments should be conducted on multilayer networks with a higher number of layers than 2 or 3 and include more complex scenarios like directed and weighted multilayer networks. Most of the papers use models based on SIS and SIR and their extensions. Studies focused on modeling single process or competing processes within single networks use much wider range of models based on random walk \cite{pu2015epidemic}, linear threshold \cite{chen2010scalable}, independent cascades \cite{kempe2005influential}, opinion formation models \cite{sznajd2000opinion} and others. It shows research gap and the area for further exploration. The situation is similar to early--stage studies focused on information spreading based on epidemiology models used before more dedicated solutions were proposed.

Also, the current landscape of research, in respect to variety of applications considered, is rather homogeneous. There is only small number of scientists who focus on problems not related to disease spreading, like products competitions or fake news. There is a need to diversify the field in this respect as well. The potential future application should include, e.g. critical infrastructures networks, financial networks or biological networks. This is a very important aspect as the bigger the variety of applications, the more we can learn about the characteristics of different processes involved. Different applications areas bring more challenges but at the same time enable to learn from each other.

Also, from the perspective of modelling, most of the parameters are set at the global level, e.g. the probability of infection, but we know that different people can respond differently to a given spread. There is a natural diversity in peoples' behaviours; ergo, there is a need for some degree of modelling at the local level.

Another element, which requires more attention, is the spread timeline. For now, the timeline conditions, for majority of analysed studies, are simplified -- all processes start propagating in the same point in time. We are all aware that it would be hard to observe such situation in real systems, but yet, there is very limited research into more realistic modelling. Modifying the timeline of different processes can strongly affect the results and obviously builds on diversity. For example, spreading awareness long time before epidemics can be crucial for suppressing actions.

We also need to start consistently including in the equation potential edges between layers and enable the processes to switch between layers. This is what happens, e.g. in information network, where spread switches between layers when information is shared from one online platform to another. Moreover, the external environment should be considered in more systematic way. Including context of the system cannot be neglected as there is no system that operates in a vacuum. 

While we are very well aware that more diverse environment increases the number of degrees of freedom of the whole problem and in consequence dramatically builds on the complexity, we are convinced that this is the only way forward as it enables to fully understand real-world spreading.

\subsection{Need for data-driven approaches}

To be able to create more realistic models, we need to depart from purely model-driven spread analysis and focus more on data-drive approaches. Few decades ago, access to data was a luxury. Currently, more and more data is available and we should take advantage of that. 

Researchers should move more towards working with real-world networks and not with networks that are small-scale, very simple, not resembling real networks and based only on theoretical models. Mobile technologies and Internet of Things create not widely explored infrastructure for spreading procesess monitoring and real data collection with the use of sensors. There is a need for representative benchmark datasets (including the real-world ones like \cite{jankowski2017multilayer}). Also, the spreading models should better reflect the real-world scenarios, e.g. by learning the values of certain parameters from data.

While this need is easy to express, it is really challenging to address. Thus, this is a whole new research area in the context of multispread over multilayer networks.

\subsection{Need for dynamic and predictive modelling}

Spreading processes are dynamic in their nature, and this has been explored by the community, but there is also a need to analyse how the spread itself influences the dynamics of the system, and how the dynamics of the system influences the spread. This is a two-way interaction and this immensely increases the diversity and, in turn, complexity of the problem. It is considered in separation from the "need for diversity" as it is a substantial topic on its own.

Almost all current approaches assume that the spreading processes interact over static networks, with \cite{gule16} being one of the exceptions. This is a very crude simplification, as the networks continuously evolve and the dynamics of this evolution changes over time. Also, the propagation process influences the dynamics of the structure itself. At the local level, currently, the dynamics of behaviour of nodes (people) is not considered, and we need to develop approaches that are able to incorporate those behavioural changes in spread models. 
How to take the node and network evolution into account when modelling the multispread is a big open question and one of the biggest challenges yet. The first steps to address the dynamics of the network could be looking into how existing spread models behave if the underlying network structure changes. The structure can be altered using certain rules, e.g. (i) random appearance/disappearance of the connextions/nodes; (ii) creation of links according to “friend-of-a-friend is my friend” phenomenon or (iii)  “rich get richer” rule. Alternative is to use real-world networks with their dynamics and overlay the spreading process on top of it. Another, more holistic approach would be to use multi-agent modelling to simulate both dynamics of and on the networks.

Element tightly connected with the dynamics is the prediction of the dynamics of spreading processes. Most of the research focuses on spread analysis and there is just limited research in the area of predictive analytics in this space \cite{weva13}, \cite{wei2012competing}. Predicting how a given spread process will behave in a given setting without running extensive simulations is an interesting avenue to explore.

\section{Conclusions}
\label{conclusions}
In this paper, we have investigated almost sixty papers about multiple spreading processes in multilayer networks. Our discussion and conclusions revolve around four fundamental questions - \textit{what spreads?}, \textit{where it spreads?}, \textit{how it spreads?} and \textit{why it spreads in that way?}. Our findings show that this is still emerging field where the number of papers and studies are increasing each year. At the same time, we found out that the research focuses only on just a few variables and is usually simplified to basic cases (only few network models and few basic spread models). 

A large fraction of papers focuses on disease spreading of multiple viruses or epidemics competing with awareness spread. Relatively low number of papers concentrates on content like memes, opinion spreading processes, decision making, and other behaviours. Research is done mainly on synthetic two layer networks where each layer is generated using some standard network model, while the processes on those layers are simulated mainly by SIS and SIR like models. We believe that it is time to depart from \textit{simple is beautiful} and start thinking that \textit{complexity is not a problem}.

The presented review offers complete and up to date view on an emerging area of multispread in multilayer networks. It identifies both contributions and drawbacks of reviewed works. We advocate that four NEEDS should drive the future research in this area: (i) a need for overarching and rigid methodology, (ii) a need for diversity, (iii) a need for data-driven approaches and (iv) a need for dynamic and predictive modelling. 

\bibliographystyle{unsrt}
\bibliography{interacting}


\EOD

\end{document}